
\magnification1200
\centerline{}
\vskip 2cm
\centerline{\bf Trees, forests and jungles:}
\medskip
\centerline {\bf a botanical garden for cluster expansions}
\vskip 2.5cm
\centerline{A. Abdesselam}
\centerline{V. Rivasseau}

\vskip .5cm

\centerline{Centre de physique th\'eorique, CNRS, UPR14}

\centerline{Ecole Polytechnique, 91128 Palaiseau Cedex, France}
\vskip 3cm

\centerline{\bf ABSTRACT}
Combinatoric formulas for cluster expansions have been improved
many times over the years. Here we develop some new combinatoric
proofs and extensions of the tree formulas of Brydges
and Kennedy, and test them on a series of pedagogical examples.
\vskip 3.5cm
\line {A.325.0994  \hfil September 1994}
\vfill\eject

\def\cN{{\cal N}}
\def\cF{{\cal F}}
\def\cV{{\cal V}}
\def\cR{{\cal R}}
\def\cD{{\cal D}}
\def\cC{{\cal C}}
\def\cO{{\cal O}}
\def\cP{{\cal P}}
\def\Pn{{{\cal P}_n}}
\def\Pnm{{{\cal P}_{n,m}}}
\def\gF{{\frak F}}
\def\gT{{\frak T}}
\def\gG{{\frak G}}
\def\toto{{\rm total}}
\def\Br{\overline}
\tolerance=10000
\hsize=17truecm\vsize=23truecm
\parindent=40pt
\mathsurround=0pt
\multiply\baselineskip by 15\divide \baselineskip by 10

%
\def\al{\alpha}

\def\de{\delta}
\def\ep{\epsilon}

\def\et{\eta}

\def\la{\lambda}

\def\ta{\tau}

\def\ph{\phi}
\def\ch{\chi}

\def\Ga{\Gamma}

\def\La{\Lambda}

\def\Ph{\Phi}

%
%
\font\tenfrak=eufm10\font\sevenfrak=eufm7\font\fivefrak=eufb5
\newfam\frakfam
     \textfont\frakfam=\tenfrak
     \scriptfont\frakfam=\sevenfrak
     \scriptscriptfont\frakfam=\fivefrak
\def\frak{\fam\frakfam\tenfrak}

\def\cA{{\cal A}}
\def\cB{{\cal B}}

\def\cF{{\cal F}}

\def\cO{{\cal O}}
\def\cS{{\cal S}}

\def\CC{\bbbc}
\def\bbbr{{\rm I\!R}}
\def\bbbc{{\mathchoice {\setbox0=\hbox{$\displaystyle\rm C$}\hbox{\hbox 
to0pt{\kern0.4\wd0\vrule height0.9\ht0\hss}\box0}}
{\setbox0=\hbox{$\textstyle\rm C$}\hbox{\hbox
to0pt{\kern0.4\wd0\vrule height0.9\ht0\hss}\box0}}
{\setbox0=\hbox{$\scriptstyle\rm C$}\hbox{\hbox
to0pt{\kern0.4\wd0\vrule height0.9\ht0\hss}\box0}}
{\setbox0=\hbox{$\scriptscriptstyle\rm C$}\hbox{\hbox
to0pt{\kern0.4\wd0\vrule height0.9\ht0\hss}\box0}}}}
\def\FF{\hbox to 8.33887pt{\rm I\hskip-1.8pt F}}
\def\NN{\hbox to 9.3111pt{\rm I\hskip-1.8pt N}}
\def\PP{\hbox to 8.61664pt{\rm I\hskip-1.8pt P}}
\def\QQ{\rlap {\raise 0.4ex \hbox{$\scriptscriptstyle |$}}
{\hskip -4.5pt Q}}
\def\RR{\hbox to 9.1722pt{\rm I\hskip-1.8pt R}}

%
%

\def\bbbone{{\mathchoice {\rm 1\mskip-4mu l} {\rm 1\mskip-4mu l}    
{\rm 1\mskip-4.5mu l} {\rm 1\mskip-5mu l}}}

%
%
\def\newenvironment#1#2#3#4{\long\def#1##1##2{\removelastskip\penalty-100
\vskip\baselineskip\noindent{#3#2\if!##1!.\else\unskip\ \ignorespaces
##1\unskip\fi\ }{#4\ignorespaces##2\vskip\baselineskip}}}
\newenvironment\lemma{Lemma}{\bf}{\it}
\newenvironment\proposition{Proposition}{\bf}{\it}
\newenvironment\theorem{Theorem}{\bf}{\it}
\newenvironment\corollary{Corollary}{\bf}{\it}
\newenvironment\example{Example}{\it}{\rm}
\newenvironment\problem{Problem}{\bf}{\rm}
\newenvironment\definition{Definition}{\bf}{\rm}
\newenvironment\remark{Remark}{\bf}{\it}

%
%
\long\def\proof#1{\removelastskip\penalty-100\vskip\baselineskip\noindent{\bf
            Proof\if!#1!\else\ \ignorespaces#1\fi:\ }\ \ \ignorespaces}
\long\def\prf{\removelastskip\penalty-100\vskip\baselineskip\noindent{\bf
            Proof:\ }\ \ \ignorespaces}
\def\sq{\hbox{\rlap{$\sqcap$}$\sqcup$}}
\def\qed{\ifmmode\sq\else{\unskip\nobreak\hfil
           \penalty50\hskip1em\null\nobreak\hfil\sq
           \parfillskip=0pt\finalhyphendemerits=0\endgraf}\fi}
\def\endproof{\hfill\vrule height .6em width .6em depth
0pt\goodbreak\vskip.25in }

\medskip
\noindent{\bf I. Introduction}
\medskip
Cluster expansions have a reputation of being hard to use; this is
largely due to the difficulty to capture them in
a single short formula. These expansions were introduced in constructive theory
by Glimm, Jaffe and Spencer [GJS1-2] and they were
improved or generalized over
the years [BF][BaF][Bat][B1]. For many years the Ecole Polytechnique was happy
using a cluster ``tree formula'' due to Brydges, Battle and Federbush
(see [R1-2] and references therein).
This formula expresses connected amplitudes more naturally
as sums over ``ordered
trees'' rather than regular trees; but a combinatoric lemma due to
Battle and Federbush, shows that the sum over all ordered trees corresponding
to a given ordinary Cayley tree has total weight 1.
A slightly disturbing
dissymmetry in this formula reveals itself in the need to order
in an arbitrary way the elements on which the cluster expansion is performed,
and in the particular r\^ole played by the first element or root of the tree.

Brydges kept digging for a truly satisfying,
still more beautiful formula, and with
Kennedy they obtained it in [BK] (see also [BY][B2]).
This formula shows a clear conceptual progress;
it does not require the use of
arbitrary choices such as an arbitrary ordering of the objects to decouple,
and the outcome can be written in a somewhat shorter form, involving
directly standard Cayley trees. Both formulas
share a positivity preserving property which is crucial in
constructive theory: the corresponding
interpolations of positive matrices remain positive. Both can be built
by iterating inductively some perturbation step. But the
first formula insists
in completing the cluster containing the ``first'' cube,
before building the next one. The second and better formula blindly
derives connections, hence all clusters grow symmetrically at the same time.
Therefore we prefer to call it a forest formula (see below).

However the original proof of this formula in [BK] relies on a differential
equation (Hamilton-Jacobi) which is perhaps not totally transparent.
The purpose of this paper is to derive a fully combinatoric
or algebraic proof of this type of formula, and to
show how to apply it to various examples chosen
for their pedagogic value. Many situations in
constructive theory in fact require several
cluster expansions on top of each other, and for this case
we derive a generalization of the
tree or forest formula, which we call the jungle formula.

We also derive a generalization that applies not only to exponentials of
two-body interaction potentials, or to perturbations of Gaussian measures.
It is a general interpolation formula we call the Taylor forest formula.

Finally we propose a formula that performs, in a single move, the succession of
a cluster and a Mayer expansion.

\bigskip
\noindent{\bf II. Two forest formulas and their combinatoric proofs}
\medskip
First let us recall the Brydges-Kennedy formula [BK] under the most
convenient setting for the following discussion.
Let $n\ge 1$ , be an integer ,
$I_n=\{1,\ldots,n\}$,
$\Pn=\bigl\{\{i,j\}/i,j\in I_n,i\ne j\bigr\}$ (the set of unordered pairs
in $I_n$).
Consider $n(n-1)/2$ elements $u_{\{ij\}}$ of a commutative Banach algebra
$\cal B$, indexed by the elements $\{ij\}$ of $\Pn$.
An element of $\Pn$ will be called a {\sl link}, a subset of $\Pn
$, a {\sl graph}. A graph $\gF=\{l_1,\ldots,l_{\tau}\}$ containing no loops,
i.e.{} no subset $\bigl\{\{i_1,i_2\},\{i_2,i_3\},\ldots,\{i_k,i_1\}\bigr\}$
with $k\ge 3$ elements,
will be called a {\sl u-forest} (unordered forest). A sequence $F=(l_1,\ldots,
l_{\tau})$ of links the range of which $\{l_1,\ldots,l_{\tau}\}$ is a u-forest,
will be called an {\sl o-forest} (ordered forest).

A u-forest is a union of disconnected trees, the supports of which are
disjoint subsets of $I_{n}$ called the
{{\sl connected components} or {\sl clusters} of $\gF$. Isolated points
also form clusters reduced to singletons, hence the total number of clusters
is $n- \ta$, where $\ta = |\gF |$.
Two points in the same cluster are said {\sl connected by $\gF$}.
\medskip
Now the Brydges-Kennedy forest formula states:
\theorem{II.1 [BK]}{
$$
\exp \biggl({\sum_{l\in \Pn} u_l}\biggr)=\sum_{\gF=\{l_1,\ldots,l_{\tau}\}
\atop{\rm u-forest}}\biggl(\prod_{\nu=1}^{\tau}\int_0^1dh_{l_{\nu}}\biggr)
\biggl(\prod_{\nu =1}^{\tau}u_{l_{\nu}}\biggr)\exp\bigl(\sum_{l\in \Pn}
h_l^{\gF}({\bf h}).u_{l}\bigr)\ ,
\eqno({\rm II.1})
$$
where the summation extends over all possible lengths
$\tau$ of $\gF$, including
$\tau=0$ hence the empty forest.
To each link of $\gF$ is attached a variable of integration $h_l$; and
$h_{\{ij\}}^{\gF}({\bf h})=\inf \bigl\{h_l,l\in L_{\gF}\{ij\}\bigr\}$ where
 $L_{\gF}\{ij\}$ is the unique path in the forest $\gF$ connecting i to j.
If no such path exists, by convention $h_{\{ij\}}^{\gF}({\bf h})=0$.}

Our second forest formula writes exactly the same, except that the definition
of the $h_{\{ij\}}^\gF(h)$ is different, and it involves for each cluster
a particular choice of a root. For simplicity let us slightly
restrict this arbitrary choice by imposing the following rule :
for each non empty subset or cluster $C$ of $I_n$,
choose $r_C$, the least element in the natural ordering
of $I_n=\{1,\ldots,n\}$, to be the root of all the trees with support $C$
that appear in the following expansion.
Now if $i$ is in some tree $\gT$ with support $C$ we call the {\sl height}
of $i$ the number of links in the unique path of the tree $\gT$ that goes from
$i$ to the root $r_C$. We denote it by $l^\gT(i)$. The set of points $i$
with a fixed height $k$ is called the $k$-th {\sl layer} of the tree.

We now have
\theorem{II.2}{
$$
\exp \biggl({\sum_{l\in \Pn} u_l}\biggr)=\sum_{\gF=\{l_1,\ldots,l_{\tau}\}
\atop{\rm u-forest}}\biggl(\prod_{\nu=1}^{\tau}\int_0^1dw_{l_{\nu}}\biggr)
\biggl(\prod_{\nu =1}^{\tau}u_{l_{\nu}}\biggr)\exp\bigl(\sum_{l\in \Pn}
w_l^{\gF}({\bf w}).u_{l}\bigr)\ ,
\eqno({\rm II.2})
$$
where the summation extends over all possible lengths
$\tau$ of $\gF$, including
$\tau=0$ hence the empty forest.
To each link of $\gF$ is attached a variable of integration $w_l$.
We define the $w_l^\gF$ as follows.
$w_{\{ij\}}^\gF({\bf w})=0$ if $i$ and $j$ are not connected by the $\gF$.
If $i$ and $j$ fall in the support $C$ of the same tree $\gT$ of $\gF$ then

$w_{\{ij\}}^\gF({\bf w})=0$\ \ if\ \ $|l^\gT(i)-l^\gT(j)|\ge2$\ \
($i$ and $j$ in distant layers)

$w_{\{ij\}}^\gF({\bf w})=1$\ \ if\ \ $l^\gT(i)=l^\gT(j)$\ \
($i$ and $j$ in the same
layer)

$w_{\{ij\}}^\gF({\bf w})=w_{\{ii'\}}$\ \ if\ \ $l^\gT(i)-1=l^\gT(j)=l^\gT(i')$,
and $\{ii'\}\in\gT$. ($i$ and $j$ in neighboring layers, $i'$ is then unique).
In particular, if $\{ij\}\in\gF$, then $w_{\{ij\}}^\gF({\bf w})=w_{\{ij\}}$.
}

The proof we give of theorem II.1 relies on two lemmas.

\lemma{II.1}{
 Let $a_0,\ldots,a_p, (p\ge1)$ be distinct complex numbers, then:
\medskip
$$
\int_{t_0\ge0,\ldots,t_p\ge0 \atop {t_0+\ldots+t_p=1}}dt_0\ldots dt_p\
\exp (t_0a_0+\ldots+t_pa_p)=\sum_{i=0}^p {{e^{a_i}} \over {\prod_{j=0 \atop
{j\ne i}}^p(a_i-a_j)}}\eqno({\rm II.3})
$$
}

\prf
By induction. $p=1$ is an easy computation.
We assume the result is true for $p\ge1$ thus
\medskip
$$
\int_{t_0\ge0,\ldots,t_{p+1}\ge0 \atop t_0+\ldots+t_{p+1}=1}dt_0\ldots dt_{p+1}
\;\exp(t_0a_0+\ldots+t_{p+1}a_{p+1})=
$$
$$
\int_0^1dt_{p+1}e^{t_{p+1}a_{p+1}}
{(1-t_{p+1})}^p\;\int_{w_0\ge0,\ldots,w_p\ge0 \atop {w_0+\ldots+w_p=1}}dw_0
\ldots dw_p\ \exp (w_0b_0+\ldots+w_pb_p)
\eqno({\rm II.4})
$$
\medskip
\noindent where we performed first the integration on $t_{p+1}$ and made the
following change of variables: $t_i=(1-t_{p+1})w_i, b_i=a_i(1-t_{p+1})$ for
$0\le i\le p$. By the induction hypothesis this becomes
$$
\int_0^1dt_{p+1}e^{t_{p+1}a_{p+1}}{(1-t_{p+1})}^p\
\sum_{i=0}^p {{e^{b_i}} \over {\prod_{j=0 \atop
{j\ne i}}^p(b_i-b_j)}}=
$$
$$
\sum_{i=0}^p {{e^{a_i}} \over {\prod_{j=0 \atop
{j\ne i}}^{p+1}(a_i-a_j)}}
+\ e^{a_{p+1}}\ \sum_{i=0}^p {{1} \over {\bigl(\prod_{{j=0} \atop {j\ne i}}^p
(a_i-a_j)\bigr)(a_{p+1}-a_i)}}\ \ .
\eqno({\rm II.5})
$$
\noindent This boils down to the wanted expression for $p+1$ after remarking
that we have the following rational fraction decomposition with simple poles:
$${1 \over {(X-a_0)\ldots(X-a_p)}}=\sum_{i=0}^p {1 \over {(X-a_i)}}\times
{1 \over {\prod_{j=0 \atop{j\ne i}}^p (a_i-a_j)}}\ \ ,
$$
\noindent and putting $X=a_{p+1}$.
\endproof
\medskip
\lemma{II.2}{
With the same notation as in the beginning of this section, assume that we are
given two sets of $n(n-1)/2$ indeterminates $u_{\{ij\}}$ and $v_{\{ij\}}$;
then the following algebraic identity is true in the field of rational
fractions $\CC(u_{\{ij\}},v_{\{ij\}})$.
$$
\prod_{l\in\Pn}v_l=\sum_{F=(l_1,\ldots,l_{\tau}) \atop {\rm o-forest}}\
u_{l_1}\ldots u_{l_{\tau}}\ .\ \Biggl( \sum_{\nu =0}^{\tau} {{b_{\nu}^F} \over
{\prod_{\mu=0 \atop \mu\ne\nu}^{\tau}(a_{\nu}^F-a_{\mu}^F)}} \Biggr)
\eqno({\rm II.6})
$$
\noindent
where, for $0\le\nu\le\tau$, $a_{\nu}^F=\sum_{\{ij\}/\nu} u_{\{ij\}}$,
$b_{\nu}^F=\prod_{\{ij\}/\nu} v_{\{ij\}}$, $\{ij\}/\nu$ meaning that the
points i and j are connected by the sub-o-forest $(l_1,\ldots,l_{\nu})$ of $F$.
In particular since no points are connected by the empty forest $a_0^F=0,
b_0^F=0.$
}

\prf
Both sides of this identity are polynomials in the $v_l$'s,
and we prove equality
coefficient by coefficient. First consider the case of the constant monomial.
We must show
$$
\sum_{F=(l_1,\ldots,l_\tau) \atop {\rm o-forest}}\ {(-1)}^\tau {{u_{l_1}\ldots
u_{l_\tau}} \over {a_1^F\ldots a_\tau^F}}
\ \ = \ \
\biggm\{
{{0\ {\textstyle if}\ n\ge2 }
\atop {1\ {\textstyle if}\ n=1} }
\eqno({\rm II.7})
$$
For n=1 it is trivial, so assume $n\ge2$ and denote $\cA_F$ the contribution
of $F$ to the left hand side. Given an o-forest $F=(l_1,\ldots,l_\tau)$,
an o-forest $F'$ of the form $(l_1,\ldots,l_\tau,l_{\tau+1},\ldots,l_{\tau
+\kappa})$ will be called a $\kappa${\sl -extension} of $F$. When we need
not know $\kappa$ we simply say an {\sl extension} of $F$.

Let\ \ \  $\cR_F=\cA_F\ .\ {\displaystyle{{a_\tau^F} \over {a_{\toto}}}}
\ \ \ {\rm where}\ \ \ a_{\toto}=\sum_{l\in\Pn} u_l \ \ \ (\ne0
\ {\rm since}\ n\ge2)$\ \ .
\medskip\noindent
Notice that
$$
a_{\toto}=a_\tau^F+\sum_{F'\ 1-{\rm extension\ of}\ F} u_{l_{\tau+1}}\ \ ,
\eqno({\rm II.8})
$$
because summing over $F'$ is the same as summing over all links $\{ij\}$ where
$i$ and $j$ lie in different clusters of $F$, while $a_\tau^F$ handles summing
over links where $i$ and $j$ are in the same cluster.

Hence
$$
a_\tau^F=a_{\toto}+\sum_{F'\ 1-{\rm extension\ of}\ F} \bigl(-{{u_{l_{\tau+1}}}
\over {a_{\tau+1}^{F'}}}\bigr){a_{\tau+1}^{F'}}
\eqno({\rm II.9})
$$

\noindent
i.e multiplying by $\cA_F/a_{\toto}$
$$
\cR_F=\cA_F+\sum_{F'\ 1-{\rm extension\ of}\ F}\cR_{F'}\ \ .
\eqno({\rm II.10})
$$
Multiple (but finite because $\tau\le n-1$) iteration of (II.10) yields
$$
\cR_F=\sum_{F'\ {\rm extension\ of}\ F}\cA_F,
\eqno({\rm II.11})
$$
where the sum is over extensions
of all possible lengths. In particular
$$\sum_F\cA_F=\sum_{F\ {\rm extension\ of}\ \emptyset}\cA_F
=\cR_{\emptyset}=\cA_{\emptyset}.{{a_0^{\emptyset}}\over{a_{\toto}}}=0
\eqno({\rm II.12})
$$
since $\cA_{\emptyset}=1,\ a_0^{\emptyset}=0,\ a_{\toto}\ne0$.
(II.7) is now proven.

Let us now check the other monomials in the $v$'s. If $F=(l_1,\ldots,l_\tau)$
is a forest we say that it {\sl creates} the partition $\cD$ of $I_n$
if $\cD$ is the set of clusters of $F$.
We then write, with a slight abuse of notation
$a_\cD=\sum_{\{ij\}/\cD}u_{\{ij\}},\ b_\cD=\prod_{\{ij\}/\cD}v_{\{ij\}}$,
where $\{ij\}/\cD$ means that i and j are in the same element or {\sl
component} of $\cD$.

Monomials generated by (II.5) are of the form $b_\cD$, $\cD$ being created by
a sub-o-forest of $F$. Remark the if $\cD$ is created by a forest of length
$\tau$ then $|\cD|=n-\tau$. There are two cases to be treated:

\noindent{\bf Case 1: $\cD\ne \{I_n\}$}

Here the coefficient of $b_\cD$ is zero in the left hand side,
we must show that also for the right hand side.
Let $\nu=n-\tau$, then (st abbreviates ``such that'')
$$
\sum_{\tau\ge\nu}\sum_{{F=(l_1,\ldots,l_\tau)} \atop {{\rm o-forest\ st}
\atop {(l_1,\ldots,l_\nu)\ {\rm creates}\ \cD}}}
{{u_{l_1}\ldots u_{l_\tau}} \over {\prod_{\mu=0 \atop \mu\ne\nu}^\tau
(a_\nu^F-a_\mu^F)}}=
$$
$$
\sum_{{F_1=(l_1,\ldots,l_\nu)} \atop {{\rm o-forest} \atop
{{\rm creating}\ \cD}}}
{{u_{l_1}\ldots u_{l_\nu}} \over {\prod_{\mu=0}^{\nu-1}(a_\cD-a_\mu^{F_1})}}
\sum_{{{F_2}=(l_{\nu+1},\ldots,l_\tau)} \atop {{{\rm st}\ F=(F_1,F_2)\
{\rm is}}
\atop {{\rm an\ o-forest}}}}{(-1)}^{\tau-\nu}{{u_{\nu+1}\ldots u_{l_\tau}}
\over{\prod_{\mu=\nu+1}^\tau (a_\mu^F-a_\cD)}}\ \ .
\eqno({\rm II.13})
$$
Again, summation is over all possible $\tau$'s.
We now arrive at the heart of the inductive argument.
We show using (II.7) that $F_1$ being fixed the sum on $F_2$ vanishes.
In fact, this sum is the analog of (II.7) when instead of
$I_n$ we use $\cD$ as a point set.
If $a$ and $b$ are two elements of the partition $\cD$, we let
${\Br u}_{\{ab\}}=\sum_{i\in a, j\in b}u_{\{ij\}}$.
Given an o-forest $F=(F_1,F_2)$ on $I_n$ such that $F_1$ creates $\cD$,
$F_2=(l_{\nu+1},\ldots,l_{\tau})$ {\sl induces} an o-forest
${\Br F}=({\Br l}_1,\ldots,{\Br l}_{\tau-\nu})$ on $\cD$ in the
following way:
if $l_\kappa=\{ij\}$, $\nu+1\le\kappa\le\mu$, with $i\in a$ and $j\in b$,
$a$ and $b$ elements of $\cD$, then set ${\Br l}_{\kappa-\nu}=\{ab\}$.
${\Br F}$ is simply obtained by forgetting the details of structure
inside the components of $\cD$, which are due to the forest $F_1$.
Demanding that the whole of $(F_1,F_2)$ be an o-forest insures that,
in the process, ${\Br F}$ remains a forest.
As a result, the analog of the $a_\nu^F$ for the forest ${\Br F}$ are
${\Br a}_\rho^{\Br F}=a_{\rho+\nu}^F-a_\cD$, for $1\le\rho\le\mu-\nu$;
and with ${\Br \tau}=\tau-\nu$ we have
$$
\sum_{{{F_2}=(l_{\nu+1},\ldots,l_\tau)} \atop {{{\rm st}\ F=(F_1,F_2)
\ {\rm is}}
\atop {{\rm an\ o-forest}}}}{(-1)}^{\tau-\nu}{{u_{\nu+1}\ldots u_{l_\tau}}
\over
{\prod_{\mu=\nu+1}^\tau (a_\mu^F-a_\cD)}}=\sum_{{\Br F}=({\Br l}_1,
\ldots,{\Br l}_{\Br \tau})}{(-1)}^{\Br \tau}{{{\Br u}_{{\Br l}_1}\ldots
{\Br u}_{{\Br l}_{\Br \tau}}} \over {{\Br a}_1^{\Br F}\ldots
{\Br a}_{\Br \tau}^{\Br F}}}\ \ .
\eqno({\rm II.14})
$$
The last sum is zero by (II.7) because $|\cD|\ge2$, since we are in
the case $\cD\ne\{I_n\}$.

\noindent{\bf Case 2: $\cD=\{I_n\}$}

Here $a_\cD=a_\toto$. Equating the coefficients of $b_\cD=\prod_{l\in\Pn}
v_l$ on both sides needs showing
$$
1=\sum_{{F=(l_1,\ldots,l_{n-1})} \atop {\rm o-forest}}{{u_{l_1}\ldots
u_{l_{n-1}}}
\over {(a_\toto-a_0^F)\ldots(a_\toto-a_{n-2}^F)}}
\eqno({\rm II.15})
$$
This last identity is shown by induction on n. Remark that here $F$ is a
complete tree covering $I_n$.
If $n=1$, the only forest is the empty one, its contribution is the empty
product i.e. 1.

The induction step form $n$ to $n+1$ needs an argument similar to the former
case:
$$
\sum_{{F=(l_1,\ldots,l_n)} \atop {{\rm o-forest\ on}\ I_n}}{{u_{l_1}\ldots
u_{l_n}}
\over {(a_\toto-a_0^F)\ldots(a_\toto-a_{n-1}^F)}}
=
$$
$$
\sum_{F_1=(l_1)}{{u_{l_1}} \over {a_\toto}}
\sum_{{F_2=(l_2,\ldots,l_n)} \atop {{{\rm st}\ F=(F_1,F_2)\ {\rm is}} \atop
{{\rm an\ o-forest\ on}\ I_{n+1}}}}{{u_{l_2}\ldots u_{l_n}}
\over {(a_\toto-a_1^F)\ldots(a_\toto-a_{n-1}^F)}}\ \ .
\eqno({\rm II.16})
$$
The link $l_1=\{\alpha,\beta\}$ being fixed, this determines a partition
$J_n=\bigl\{\{\alpha,\beta\}\bigr\}\cup \bigl\{\{i\}/i\in I_{n+1},\ i\ne\alpha,
\ i\ne\beta\bigr\}$
for which we can repeat the treatment of the partition $\cD$ in the preceding
case:
introduce variables ${\Br u}_{\{ab\}}=\sum_{i\in a,\ j\in b}u_{\{ij\}}$ for
$a$ and $b$ in $J_n$, and induced o-forests ${\Br F}=({\Br l}_1,\ldots,
{\Br l}_{n-1})$ on $J_n$ for any o-forest $F_2=(l_2,\ldots,l_n)$ on $I_n$
such that $F=(F_1,F_2)$ be an o-forest on $I_{n+1}$.
Remark that
$$
{\Br a}_\toto{{\rm def} \atop {=\atop {\ }}}
\sum_{{\{ab\}\subset J_n} \atop {a\ne b}}
{\Br u}_{\{ab\}}
=\biggl(\sum_{l\in {\cal P}_{n+1}}u_l\biggr)-u_{\{\alpha\beta\}}=a_\toto-a_1^F
\ \ ,
\eqno({\rm II.17})
$$
and for $0\le\rho\le n-2$, ${\Br a}_\rho^{\Br F}=a_{\rho+1}^F-a_1^F$,
therefore ${\Br a}_\toto-{\Br a}_\rho^{\Br F}=(a_\toto-a_1^F)-(a_{\rho+1}^F
-a_1)=a_\toto-a_{\rho+1}^F$.

Hence
$$
\sum_{{F_2=(l_2,\ldots,l_n)} \atop {{{\rm st}\ F=(F_1,F_2)\ {\rm is}} \atop
{{\rm an\ o-forest\ on}\ I_{n+1}}}}{{u_{l_2}\ldots u_{l_n}}
\over {(a_\toto-a_1^F)\ldots(a_\toto-a_{n-1}^F)}}
=
$$
$$
\sum_{{{\Br F}=({\Br l}_1,\ldots,{\Br l}_{n-1})} \atop
{{\rm o-forest\ on}\ J_n}}
{{{\Br u}_{{\Br l}_1}\ldots{\Br u}_{{\Br l}_{n-1}}} \over
{({\Br a}_\toto-{\Br a}_0^{\Br F})\ldots({\Br a}_\toto-{\Br a}_{n-2}^{\Br F})}}
\eqno({\rm II.18})
$$
\noindent but the last sum is 1 by the induction hypothesis since $|J_n|=n$.
Finally in (II.15) there remains
$$
\sum_{l_1}{{u_{l_1}} \over {a_\toto}}={{a_\toto} \over {a_\toto}}=1\ \ .
\eqno({\rm II.19})
$$
This completes the proof of Lemma II.2 .
\endproof
\medskip
\noindent {\bf Proof of theorem II.1:}

We can return now to the Brydges-Kennedy forest formula and first prove it for
any complex numbers $u_l$, $l\in\Pn$ such that for all subset $X$ of $\Pn$,
$\sum_{l\in X}u_l\ne 0$. In fact we can rewrite the right hand side of (II.1)
using o-forests instead of u-forests. Let $\gF$ be a u-forest and choose an
ordering $F=(l_1,\ldots,i_\tau)$ of its elements. We can slice the integral
according to the ordering of the $h_l$'s thereby giving a contribution
for $\gF$
$$
\sum_{\sigma\in{\frak S}_\tau}
\int_{1>h_{l_{\sigma(1)}}>\ldots>h_{l_{\sigma(\tau)}}>0}
\biggl(\prod_{\nu=1}^\tau dh_{l_\nu}\biggr)
\biggl(\prod_{\nu =1}^{\tau}u_{l_{\nu}}\biggr)\exp\bigl(\sum_{l\in \Pn}
h_l^{\gF}({\bf h}).u_{l}\bigr)\ \ ,
\eqno({\rm II.20})$$
${\frak S}_\tau$ is the permutation group on $\tau$ elements.
For any of its elements $\sigma$ let us denote the o-forest
$(l_{\sigma(1)},\ldots,l_{\sigma(\tau)})$ by $F_\sigma$.
Let us define, if $F$ is an o-forest, the function
$h_{\{ij\}}^F=h_{l_\nu}$
where $\nu$ is  the largest index in the ordering provided by F of the
links appearing in the unique path of F connecting $i$ and $j$.
If no such path exists we again put by convention $h_{\{ij\}}^F=0$.

Thus the contribution of $\gF$ becomes
$$
\sum_{\sigma\in{\frak S}_\tau}
\int_{1>h_{l_{\sigma(1)}}>\ldots>h_{l_{\sigma(\tau)}}>0}
\biggl(\prod_{\nu=1}^\tau dh_{l_\nu}\biggr)
\biggl(\prod_{\nu =1}^{\tau}u_{l_{\nu}}\biggr)\exp\bigl(\sum_{l\in \Pn}
h_l^{F_\sigma}({\bf h}).u_{l}\bigr)
\eqno({\rm II.21})
$$
and the right hand side of (II.1)
$$
I=
\sum_{{F=(l_1,\ldots,l_\tau)} \atop {\rm o-forest}}
\int_{1>h_{l_1}>\ldots>h_{l_\tau}>0}
\biggl(\prod_{\nu=1}^\tau dh_{l_\nu}\biggr)
\biggl(\prod_{\nu =1}^{\tau}u_{l_{\nu}}\biggr)\exp\bigl(\sum_{l\in \Pn}
h_l^F({\bf h}).u_{l}\bigr)\ .
\eqno({\rm II.22})
$$

Now we perform the change of variables $t_1=h_{l_1}-h_{l_2},\ldots,
t_{\tau-1}=h_{l_{\tau-1}}-h_{l_\tau},\ t_\tau=h_{l_\tau}$ then
$$
I=\sum_{{F=(l_1,\ldots,l_\tau)} \atop {\rm o-forest}}
u_{l_1}\ldots u_{l_\tau}
\int_{{t_1\ge0,\ldots,t_\tau\ge0} \atop {t_1+\ldots+t_\tau\le1}}
dt_1\ldots dt_\tau\; \exp \bigl(\sum_{\nu=1}^\tau t_\nu a_\nu^F\bigr)\ \ .
\eqno({\rm II.23})
$$
There shows up the $a_\nu^F$ when collecting in the exponential the
$u_{\{ij\}}$'s multiplied by a given $t_\nu$. Since $s_\nu=t_\nu+t_{\nu+1}+
\ldots+t_\tau$, the corresponding pairs $\{ij\}$ are those for which all the
links appearing in the unique path of $F$ connecting $i$ and $j$ have an
index at most equal to $\nu$, i.e. $\{ij\}/\nu$.
Since $a_0^F=0$, we can rewrite (II.23) as
$$
I=\sum_{{F=(l_1,\ldots,l_\tau)} \atop {\rm o-forest}}
u_{l_1}\ldots u_{l_\tau}
\int_{{t_0\ge0,\ldots,t_\tau\ge0} \atop {t_0+\ldots+t_\tau=1}}
dt_0\ldots dt_\tau\; \exp \bigl(\sum_{\nu=0}^\tau t_\nu a_\nu^F\bigr)\ \ ,
\eqno({\rm II.24})
$$
then from lemma II.1 we obtain
$$
I=\sum_{{F=(l_1,\ldots,l_{\tau})} \atop {\rm o-forest}}\
u_{l_1}\ldots u_{l_{\tau}}\ .\
\Biggl( \sum_{\nu =0}^{\tau} {{e^{a_\nu^F}} \over
{\prod_{\mu=0 \atop \mu\ne\nu}^{\tau}(a_{\nu}^F-a_{\mu}^F)}} \Biggr)\ \ .
\eqno({\rm II.25})
$$
\indent
Then if we substitute for the indeterminates $u_{l}$ the complex
numbers we have now (this is allowed since the factors in the denominators
$a_\nu^F-a_\mu^F$ are of the form $\pm\sum_{l\in X}u_l\ne0$) and for
the $v_l$
the complex numbers $\exp(u_l)$, the lemma II.2 just proves (II.1).
We can remove the condition
$\sum_{l\in X}u_l\ne 0$ of being in a finite number of hyperplanes
of $\CC^\Pn$,
by density and continuity of both sides of
(II.1). To prove it for an arbitrary commutative Banach algebra ${\cal B}$
we need only note that the identity we are about to show, is of the form
$f(u_1,\ldots,u_k)=0$, where the function $f$ is analytic, its value for
$u$'s in a commutative Banach algebra being defined by its power series
expansion. Thus proving it for all complex numbers is enough to entail
the vanishing of the coefficients of the power series and to prove the identity
for any $u$'s in ${\cal B}$.
\endproof
\medskip
\noindent {\bf Proof of theorem II.2:}

We recall the notion of connected set function.
If $X$ is a finite set and $\psi$ is a map from $\cP(X)$, the power set of $X$,
to a commutative algebra $\cal B$, the connected function of $\psi$ is
$\psi_c:\cP(X)\to{\cal B}$ defined inductively by
$$
\forall C\subset X,\ \ \psi(C)=
\sum_{{\Pi=\{C_1,\ldots,C_k\}} \atop {{\rm partition\ of}\ C}}
\prod_{i=1}^k\psi_c(C_i)
\ \  ,\ \ \psi_c(\emptyset)=1\ \ .
\eqno({\rm II.26})
$$
Let $X=I_n$, $\cal B$ be the field of complex numbers,
$\forall l\in\Pn, \epsilon_l{{\rm def}\atop{=\atop{ }}}e^{u_l}-1$, and let
$\psi(C)=\prod_{\{ij\}\subset C}(1+\epsilon_{\{ij\}})$.
We make also the assumption that for any non empty subset $Q$ of $\Pn$,
$\sum_{l\in Q}u_l\ne 0\;$.
It is well known [B1] that
$$
\psi_c(C)=\sum_{\gG\ {\rm graph\ connecting}\ C}(\prod_{l\in\gG}\epsilon_l)
\eqno({\rm II.27})
$$
\noindent
(``connecting $C$'' means that for every $i\ne j$ in $C$ there exists a path
in $\gG$ going from $i$ to $j$, the empty graph connects $C$ if $|C|\le1$).

We claim that
$$
\psi_c(C)=
\sum_{\gT\ {\rm tree\ connecting}\ C}
\biggl(\prod_{l\in\gT}\int_0^1dw_l\biggr)
\biggl(\prod_{l\in\gT}u_l\biggr)
\exp\biggl(\sum_{\{ij\}\subset C}w_{\{ij\}}^\gT({\bf w}).u_{\{ij\}}\biggr)
\ \ .
\eqno({\rm II.28})
$$
This is trivial if $|C|\le1$. If $|C|\ge2$, let $\gT$ be a non trivial tree
connecting $C$. We associate the sequence $\cC^\gT={(C_\nu^\gT)}_{\nu\in\NN}$
of disjoint subsets of $C$ such that $C_\nu^\gT$
is the $\nu$-th layer of $\gT$.
Note that $C_0^\gT=\{r_C\}$, $\cup_{\nu\in\NN}C_\nu^\gT=C$, and
$\exists\nu_\gT\in\NN$ such that $\nu\le\nu_\gT\ \Rightarrow\ C_\nu^\gT\ne
\emptyset$ and $\nu>\nu_\gT\ \Rightarrow\ C_\nu^\gT=\emptyset$;
a sequence $\cC={(C_\nu)}_{\nu\in\NN}$ sharing those three properties
will be called {\sl admissible}.

We will compute in the right hand side of (II.28) the contribution of the trees
corresponding to a given sequence $\cC^\gT=\cC$.
Let $\{ij\}$ be a pair in $C$ with $i\in C_{\nu_i}$, $j\in C_{\nu_j}$ and
$\nu_i\ge\nu_j$. The only way $w_{\{ij\}}^\gT({\bf w})$ can be non zero is
that $\nu_i=\nu_j$ or $\nu_i=\nu_j+1$.
For $\nu_i=\nu_j$ we just get a factor $e^{u_{\{ij\}}}=1+\epsilon_{\{ij\}}$,
since $w_{\{ij\}}^\gT({\bf w})=1$.

But if $\nu_i=\nu_j+1$, let $i'$ be the element of $C_{\nu_j}$ such that
$\{ii'\}\in\gT$; $i'$ is the {\sl ancestor} of $i$. Let us perform the
integration explicitly in $w_{\{ii'\}}$. The pairs $l$ such that
$w_l^\gT({\bf w}){{\rm def}\atop{=\atop{ }}}w_{\{ii'\}}$ are the pairs
$\{ij\}$ with $j\in C_{\nu_i-1}$, the corresponding contribution is
$$
\int_0^1dw_{\{ii'\}}\ u_{\{ii'\}}.\exp\biggl(w_{\{ii'\}}.
\sum_{j\in C_{\nu_i-1}}
u_{\{ij\}}\biggr)=
u_{\{ii'\}}\ .\ {{\exp\Bigl(\sum_{j\in C_{\nu_i-1}}u_{\{ij\}}\Bigr)-1}
\over {\sum_{j\in C_{\nu_i-1}}u_{\{ij\}}}}\ \ .
\eqno({\rm II.29})
$$
\indent
Now, building a tree with prescribed $\cC^\gT$ means linking the points of
$C_1^\gT$ to the root $r_C$, then for each $i$ in $C_2^\gT$ choosing an
ancestor $i'$ in $C_1^\gT$, then for each $i$ in $C_3^\gT$ choosing an
ancestor $i'$ in $C_2^\gT$, and so on, until we exhaust $C$.
The choice of ancestor for points in the same layer $C_\nu^\gT$ are completely
independent operations. Summing the contributions (II.29) for all choices of
ancestors for $i$ is simply
$$
\sum_{i'\in C_{\nu_i-1}}u_{\{ii'\}}\ .\
{{\exp\Bigl(\sum_{j\in C_{\nu_i-1}}u_{\{ij\}}\Bigr)-1}
\over {\sum_{j\in C_{\nu_i-1}}u_{\{ij\}}}}
=\exp\Bigl(\sum_{j\in C_{\nu_i-1}}u_{\{ij\}}\Bigr)-1
$$
$$
=\prod_{j\in C_{\nu_i-1}}(1+\epsilon_{\{ij\}})\ -1
=\sum_{{J\subset C_{\nu_i-1}} \atop {J\ne\emptyset}}
\Bigl(\prod_{j\in J}\epsilon_{\{ij\}}\Bigr)\ \ .
\eqno({\rm II.30})
$$
Therefore the right hand side of (II.28) is
$$
\sum_{{\rm admissible}\atop{\rm sequence\ \cC}}\prod_{\nu\ge1}
\Biggl(\biggl(\prod_{i\in C_\nu}
 \Bigl(\sum_{{J\subset C_{\nu-1}}\atop{J\ne\emptyset}}\prod_{j\in J}
   \epsilon_{\{ij\}}\Bigr)\biggr)
\times
 \Bigl(\prod_{\{ij\}\subset C_\nu}(1+\epsilon_{\{ij\}})\Bigr)
\Biggr)\ \ ,
\eqno({\rm II.31})
$$
and is to be compared with (II.27).

Note that, in a connected graph $\gG$, we can define the distance $d_\gG(i,j)$
between two points $i$ and $j$ by the minimal number of links in a path in
$\gG$ going from $i$ to $j$.
Given a preferred point in $C$, namely the root $r_C$, and a graph $\gG$
connecting $C$,
we can define an admissible
sequence $\cC^\gG={(C_\nu^\gG)}_{\nu\in\NN}$, by letting
$C_\nu^\gG{{\rm def}\atop{=\atop{ }}}\{i\in C/d_\gG(i,r_C)=\nu\}$.
This is consistent with
our previous definition since a tree is obviously a special case of graph.

We just have now to convince the reader that for a given admissible $\cC$
$$
\prod_{\nu\ge1}
\Biggl(\biggl(\prod_{i\in C_\nu}
 \Bigl(\sum_{{J\subset C_{\nu-1}}\atop{J\ne\emptyset}}\prod_{j\in J}
   \epsilon_{\{ij\}}\Bigr)\biggr)
\times
 \Bigl(\prod_{\{ij\}\subset C_\nu}(1+\epsilon_{\{ij\}})\Bigr)
\Biggr)
=
\sum_{{\gG\ {\rm graph}\atop{{\rm connecting}\ C}}
\atop{{\rm such \ that}\ \cC^\gG=\cC}}
\prod_{l\in\gG}\epsilon_l
\ \ .
\eqno({\rm II.32})
$$
Note that building a graph $\gG$ with $\cC^\gG=\cC$ is first linking the
points $i$ of $C_1$ to the root $r_C$, then for each $i$ in $C_2$ choosing
a {\sl non empty} subset $J_i\subset C_1$ of points $j$ to which $i$ will be
linked, then for $i$ in $C_3$ we again choose a non empty subset
$J_i\subset C_2$ of points $j$ to be linked with, and so on. Having built
such a ``skeleton'' we, as a final step,
have complete freedom to add all the links $\{ij\}$,
for $i$ and $j$ in the same $C_\nu$, that we want.
Remark that in such a graph $\gG$, there is no link $\{ij\}$ with
$i\in C_{\nu_i}$, $j\in C_{\nu_j}$ and $\nu_j+1<\nu_i$, for we would
have $d_\gG(r_C,i)\le d_\gG(r_C,j)+1=\nu_j+1<\nu_i=d_\gG(r_c,i)$, a
contradiction.

Now (II.28) is proven, so the right hand side of (II.2) is
$$
\sum_{{\Pi=\{C_1,\ldots,C_k\}} \atop {{\rm partition\ of}\ I_n}}
\prod_{i=1}^k\psi_c(C_i)=\psi(I_n)=\prod_{l\in\Pn}(1+\epsilon_l)
=\exp\Bigl(\sum_{l\in\Pn}u_l\Bigr)\ \ .
\eqno({\rm II.33})
$$
Finally we can remove the conditions $\sum_{l\in Q}u_l\ne0$ and take the
$u_l$ in any commutative Banach algebra $\cal B$, as we did for the proof
of theorem II.1.
\endproof

\bigskip
\noindent{\bf III. A first Generalization: The Taylor forest formulas}
\medskip
In practice we not only need ``algebraic'' cluster expansions, but
also ``Taylor'' cluster expansion, in which
one typically makes a Taylor expansion with integral
remainder to interpolate between a coupled situation (parameters set to 1)
and an uncoupled one (parameters set to 0).

For instance a ``pair of cubes'' cluster expansion
[R1] boils down to applying the
following interpolation formula:
$$
f(1,\ldots,1)=\sum_{I\subset I_n}\biggl(\prod_{i\in I}\int_0^1 dh_i\biggr)
\Biggl(\biggl(\prod_{i\in I}{\partial \over {\partial x_i}}\biggr)f\Biggr)
\bigl(\psi_I({\bf h})\bigr)
\eqno({\rm III.1})
$$
\noindent where $f(x_1,\ldots,x_n)$ is a smooth function on $\RR^n$, and
$\psi_I({\bf h})$ is the vector of coordinates $(x_1,\ldots,x_n)$ defined
by $x_i=0$ if $i\not\in I$ and $x_i=h_i$ if $i\in I$.
Such a formula is easy to prove.

The ``Taylor'' tree cluster expansion describes a similar interpolation
formula.
Let $\cS$ be the space of smooth functions from
$\RR^\Pn$ to an arbitrary Banach space $\cV$.
An element of $\RR^\Pn$ will be generally denoted by ${\bf x}={(x_l)}_
{l\in\Pn}$. The vector with all entries equal to $1$ will be denoted by
$\bbbone$. Applied to an element $H$ of $\cS$, we can state two different
Taylor forest formulas depending on which of theorem II.1 or II.2 we
use for its derivation.

\theorem{III.1 (The Brydges-Kennedy Taylor forest formula)}{
$$
H(\bbbone)=
\sum_{\gF\ {\rm u-forest}}
\biggl(\prod_{l\in\gF}\int_0^1 dh_l\biggr)
\biggl(\Bigl(\prod_{l\in\gF}
{\partial \over {\partial x_l}}\Bigr)H\biggr)
\bigl(X_\gF^{BK}({\bf h})\bigr) \ \ .
\eqno({\rm III.2})
$$
\noindent Here $X_\gF^{BK}({\bf h})$ is the vector ${(x_l)}_{l\in\Pn}$
of $\RR^\Pn$ defined by $x_l=h_l^\gF({\bf h})$, which is the value
at which we evaluate the
complicated derivative of $H$.}
and
\theorem{III.2 (The rooted Taylor forest formula)}{
$$
H(\bbbone)=
\sum_{\gF\ {\rm u-forest}}
\biggl(\prod_{l\in\gF}\int_0^1 dw_l\biggr)
\biggl(\Bigl(\prod_{l\in\gF}
{\partial \over {\partial x_l}}\Bigr)H\biggr)
\bigl(X_\gF^r({\bf w})\bigr) \ \ .
\eqno({\rm III.3})
$$
\noindent Here $X_\gF^r({\bf w})$ is the vector ${(x_l)}_{l\in\Pn}$
of $\RR^\Pn$ defined by $x_l=w_l^\gF({\bf w})$, which is the value
at which we evaluate the
complicated derivative of $H$.}

\prf
These formulas look more general than the algebraic forest formulas
(II.1) and (II.2) (obtained
by letting $H({\bf x})=\exp(\sum_{l\in\Pn}x_lu_l)$ ) but are in fact a
consequence of them.
For any integer $p\ge0$, let $\cS_p$ be the space of polynomial functions in
the $x_l$'s of total degree at most $p$. $\cS_p$ can be dressed in a
finite dimensional Banach space structure. Let $\cB_p$ be the algebra of
linear operators on $\cS_p$ generated by the derivations $\partial/\partial
x_l$ for $l\in\Pn$, equipped with the operator norm. $\cB_p$ is a
finite dimensional {\sl commutative} Banach algebra. If we consider the
$u_l{{\rm def} \atop {=\atop{ }}} \partial/\partial x_l$ in $\cB_p$ we have
the right to use the multiple forest formula of the preceding section.

We then remark that, in $\cS_p$, $\exp(\lambda\partial/\partial x_l)$
is the translation operator by the vector $-\lambda e_l$, where
${(e_l)}_{l\in\Pn}$ is the canonical base of $\RR^\Pn$ and the
exponential is defined by its power series, which converges because
$\partial/\partial x_l$ is a {\sl bounded} operator on $\cS_p$. It reduces
to a translation simply because Taylor's formula is finite and exact for
polynomials.

Since $\exp(\sum_{l\in\Pn}\lambda_l\partial/\partial x_l)$
is the translation by ${\bf \lambda}=-\sum_{l\in\Pn}\lambda_l
e_l$ i.e. the operator on $\cS_p$ which maps the polynomial function
$H$ to $H':{\bf x}\mapsto H({\bf x}-{\bf \lambda})$, we have proven (III.2)
for any $H$ in $\cS_p$. But $p$ is arbitrary so we have proven the formula
for any polynomial function $H$. By density of the polynomials in the Frechet
space $\cS$ for the $C^\infty$ topology and the continuity relatively to
that topology of both sides of (III.2), we extend the validity of the equation
to all $H$ in $\cS$.
We have proved formula (III.2) but the argument is the same for formula
(III.3), just replace ``h'' by ``w''.
\endproof
\bigskip
\noindent{\bf IV. A second generalization: The jungle formulas}
\medskip
Let $m\ge1$ be an integer, and ${(u_l^k)}_{l\in\Pn}$, $1\le k\le m$,
be $m$ families of $n(n-1)/2$ elements of a commutative Banach algebra
${\cal B}$. An {\sl m-jungle} is a sequence $\cF=(\gF_1,\ldots,\gF_m)$
of u-forests on $I_n$ such that $\gF_1\subset\ldots\subset\gF_m$.

The multiple forest formula states:

\theorem{IV.1 (The algebraic Brydges-Kennedy jungle formula)}{
$$
\exp\biggl(\sum_{{l\in\Pn} \atop {1\le k\le m}} u_l^k\biggr)=
\sum_{{\cF=(\gF_1,\ldots,\gF_m)} \atop {m-{\rm jungle}}}
\biggl(\prod_{l\in\gF_m}\int_0^1 dh_l\biggr)
\biggl(\prod_{k=1}^m\Bigl(\prod_{l\in\gF_k\backslash
\gF_{k-1}}u_l^k\Bigr)\biggr)
$$
$$\ . \
\exp\biggl(\sum_{k=1}^m\sum_{l\in\Pn}h_l^{\cF,k}({\bf h}).u_l^k\biggr)
\eqno({\rm IV.1})
$$
where $\gF_0=0$ by convention, ${\bf h}$ is the vector ${(h_l)}_{l\in\gF_m}$
and the functions $h_{\{ij\}}^{\cF,k}({\bf h})$ are defined in the
following manner:

- If $i$ and $j$ are not connected by $\gF_k$ let
$h_{\{ij\}}^{\cF,k}({\bf h})=0$.

- If $i$ and $j$ are connected by $\gF_k$ but not by $\gF_{k-1}$ let
$$
h_{\{ij\}}^{\cF,k}({\bf h})=\inf\Bigl\{h_l, l\in L_{\gF_k}\{ij\}\cap(\gF_k
\backslash\gF_{k-1})\Bigr\}
$$
(recall
that $L_{\gF}\{ij\}$ is the unique path in the forest $\gF$ connecting i to j).

- If $i$ and $j$ are connected by $\gF_{k-1}$ let
 $h_{\{ij\}}^{\cF,k}({\bf h})=1$.
}

\prf
By induction. The case $m=1$ was treated in Section II.
For the induction step from $m$ to $m+1$, we sum over the last forest
$\gF_{m+1}$:
$$
\sum_{{\cF=(\gF_1,\ldots,\gF_{m+1})} \atop {(m+1)-{\rm jungle}}}
\biggl(\prod_{l\in\gF_{m+1}}\int_0^1 dh_l\biggr)
\Biggl(\prod_{k=1}^{m+1}\biggl(\prod_{l\in\gF_k\backslash\gF_{k-1}}
u_l^k\biggr)\Biggr)
\exp\biggl(\sum_{k=1}^{m+1}\sum_{l\in\Pn}h_l^{\cF,k}({\bf h}).u_l^k\biggr)
$$
$$
=\sum_{{\cF'=(\gF_1,\ldots,\gF_m)} \atop {m-{\rm jungle}}}
\biggl(\prod_{l\in\gF_m}\int_0^1 dh_l\biggr)
\Biggl(\prod_{k=1}^m\biggl(\prod_{l\in\gF_k\backslash\gF_{k-1}}
u_l^k\biggr)\Biggr)
\exp\biggl(\sum_{k=1}^m\sum_{l\in\Pn}h_l^{\cF',k}({\bf h'}).u_l^k\biggr)
$$
$$
.\ \sum_{{\gF_{m+1}\ {\rm u-forest}} \atop {\gF_m\subset\gF_{m+1}}}
\biggl(\prod_{l\in\gF_{m+1}\backslash\gF_m}\int_0^1 dh_l\biggr)
\biggl(\prod_{l\in\gF_{m+1}\backslash\gF_m}u_l^{m+1}\biggr)
\exp\biggl(\sum_{l\in\Pn}h_l^{\cF,m+1}({\bf h}).u_l^{m+1}\biggr)
\eqno({\rm IV.2})
$$
\noindent where ${\bf h}={(h_l)}_{l\in\gF_{m+1}}$, ${\bf h'}={(h_l)}_
{l\in\gF_m}$
and we have noted that if $1\le k\le m$ then $h_l^{\cF,k}({\bf h})=
h_l^{\cF',k}({\bf h'})$.
To perform the summation over $\gF_{m+1}$, we will use the forest formula of
section II and our favorite argument of
forgetting the details of the tree
structure up to $\gF_m$, to concentrate on what $\gF_{m+1}$ brings as new
connections between the existing clusters.
We introduce the partition $\cD$ of $I_n$ created by $\gF_m=\{l_1,\ldots,
l_\nu\}$ and the u-forest on $\cD$, ${\Br \gF}_{m+1}=\{{\Br l}_1,\ldots,
{\Br l}_{\tau-\nu}\}$ induced by $\gF_{m+1}\backslash\gF_m=\{l_{\nu+1},\ldots,
l_\tau\}$ with $\nu\le\tau$. The definitions are the same as in the proof
of Lemma II.2 except that we have u-forests instead of o-forests.

For a link $\{ab\}$ between two elements $a$ and $b$ of $\cD$, let
${\Br u}_{\{ab\}}=\sum_{i\in a,j\in b}u_{\{ij\}}^{m+1}$. Summing over
$\gF_{m+1}$, u-forest on $I_n$ containing $\gF_m$, with the ``propagators''
$u_{\{ij\}}^{m+1}$ is the same as summing over the u-forests ${\Br\gF}_{m+1}$
on $\cD$ with the ``propagators'' ${\Br u}_{\{ab\}}$ i.e.
$$
\sum_{{\Br\gF}_{m+1}\ {\rm u-forest\ on}\ \cD}
\biggl(\prod_{{\Br l}\in{\Br \gF}_{m+1}}\int_0^1 d{\Br h}_{\Br l}\biggr)
\biggl(\prod_{{\Br l}\in{\Br\gF}_{m+1}} {\Br u}_{\Br l}\biggr)
\exp\Biggl(\sum_{{\{ab\}\subset\cD} \atop {a\ne b}}
{\Br h}_{\{ab\}}^{{\Br\gF}_{m+1}}({\bf \Br h}).{\Br u}_{\{ab\}}\Biggr)
$$
$$
=\sum_{{\gF_{m+1}\ {\rm u-forest\ on}\ I_n} \atop {{\rm such \ that}\ \gF_m
\subset\gF_{m+1}}}
\biggl(\prod_{l\in\gF_{m+1}\backslash\gF_m}\int_0^1 dh_l\biggr)
\biggl(\prod_{l\in\gF_{m+1}\backslash\gF_m}u_l^{m+1}\biggr)
\exp\Biggl(\sum_{{\{ij\}\in\Pn} \atop {\{ij\}\sharp \cD}}h_{\{ij\}}^{\cF,m+1}
({\bf h}).u_{\{ij\}}^{m+1}\Biggr)
\eqno({\rm IV.3})
$$
\noindent where $\{ij\}\sharp \cD$ means ``$i$ and $j$ are in different
components of $\cD$''.
By the forest formula of
section II, the left hand side of the last equality
is
$$
\exp\Biggl(\sum_{{\{ab\}\subset\cD} \atop {a\ne b}}{\Br u}_{\{ab\}}\Biggr)
\eqno({\rm IV.4})$$
\noindent that is
$$
\exp\Biggl(\sum_{{\{ij\}\in\Pn} \atop {\{ij\}\sharp\cD}}
u_{\{ij\}}^{m+1}\Biggr)\ \ .
\eqno({\rm IV.5})$$
\indent
The right hand side is almost the partial sum we want to perform on
$\gF_{m+1}$ in (IV.2), there misses
$$
\exp\Biggl(\sum_{{\{ij\}\in\Pn} \atop {\{ij\}/\cD}}h_{\{ij\}}^{\cF,m+1}
({\bf h}).u_{\{ij\}}^{m+1}\Biggr)
\ \ .
\eqno({\rm IV.6})$$
The sum is over all pairs $\{ij\}$ in $I_n$ such that $\{ij\}/\cD$ i.e. $i$
and $j$ are connected by $\gF_m$. But then the definition of the functions
$h_l^{\cF,k}({\bf h})$ tells us $h_{\{ij\}}^{\cF,m+1}({\bf h})=1$; so the
missing factor becomes
$$
\exp\Biggl(\sum_{{\{ij\}\in\Pn} \atop {\{ij\}\sharp\cD}}u_{\{ij\}}^{m+1}\Biggr)
\ \ .
\eqno({\rm IV.7})
$$
\indent
In conclusion
$$
\sum_{{\gF_{m+1}\ {\rm u-forest\ on}\ I_n} \atop
{{\rm st}\ \gF_m\subset\gF_{m+1}}}
\biggl(\prod_{l\in\gF_{m+1}\backslash\gF_m}\int_0^1 dh_l\biggr)
\biggl(\prod_{l\in\gF_{m+1}\backslash\gF_m}u_l^{m+1}\biggr)
\exp\biggl(\sum_{l\in\Pn}h_l^{\cF,m+1}({\bf h}).u_l^{m+1}\biggr)
$$
$$
=\ \exp\biggl(\sum_{l\in\Pn}u_l^{m+1}\biggr)
\ \ . \eqno({\rm IV.8})
$$
The sum over the $m$-jungle $\cF'$ is now $\exp(\sum_{k=1}^m\sum_{l\in\Pn}
u_l^k)$ by the induction hypothesis, and the right hand side of (IV.2) is
finally summed to $\exp(\sum_{k=1}^{m+1}\sum_{l\in\Pn}u_l^k)$ as wanted.
\endproof
\bigskip

We can generalize in a similar way the rooted forest formula.
Given an $m$-jungle $\cF=(\gF_1,\ldots,\gF_m)$, we introduce the notation
${\bf w}$ for the vector ${(w_l)}_{l\in\gF_m}$, and $w_{\{ij\}}^{\cF,k}
({\bf w})$ for the functions defined by:

- If $i$ and $j$ are not connected by $\gF_k$ let
$w_{\{ij\}}^{\cF,k}({\bf w})=0$.

- If $i$ and $j$ are connected by $\gF_{k-1}$ let
$w_{\{ij\}}^{\cF,k}({\bf w})=1$.

- If $i$ and $j$ are connected by $\gF_k$ but not by $\gF_{k-1}$,
let $r_C$ be the root of the cluster of $\gF_k$ containing $i$ and $j$,
define $l_k(i)$ to be the number of links of $\gF_k\backslash\gF_{k-1}$
in the unique path that goes from $i$  to $r_C$, and similarly for $j$.
Now if $|l_k(i)-l_k(j)|\ge2$ put $w_{\{ij\}}^{\cF,k}({\bf w})=0$,
if $l_k(i)=l_k(j)$ put $w_{\{ij\}}^{\cF,k}({\bf w})=1$,
and if $l_k(i)=l_k(j)+1$ for instance, take $w_{\{ij\}}^{\cF,k}({\bf w})=w_l$
where $l$ is the first link of $\gF_k\backslash\gF_{k-1}$ on the path that
goes from $i$ to $r_C$. It is about the same definition as for theorem II.2
but we take only into account the links in $\gF_k\backslash\gF_{k-1}$.

We can now state

\theorem{IV.2 (The algebraic rooted jungle formula)}{

$$
\exp\biggl(\sum_{{l\in\Pn} \atop {1\le k\le m}} u_l^k\biggr)=
\sum_{{\cF=(\gF_1,\ldots,\gF_m)} \atop {m-{\rm jungle}}}
\biggl(\prod_{l\in\gF_m}\int_0^1 dw_l\biggr)
\biggl(\prod_{k=1}^m\Bigl(\prod_{l\in\gF_k\backslash
\gF_{k-1}}u_l^k\Bigr)\biggr)
$$
$$\ . \
\exp\biggl(\sum_{k=1}^m\sum_{l\in\Pn}w_l^{\cF,k}({\bf w}).u_l^k\biggr) \  .
\eqno({\rm IV.9})
$$
}
\prf
The deduction of the jungle formula from the forest formula is the same as in
the Brydges-Kennedy case. Remark that the property that was used was the
following.
Once $\gF_m$ is fixed as well as the partition $\cD$ of $I_n$ it creates,
the numbers $w_{\{ij\}}^{\cF,m+1}({\bf w})$ are unchanged if we allow $i$
to travel freely in the component $a_i$ of $\cD$ it belongs to,
and the same for $j$ in $a_j$.
Furthermore $w_{\{ij\}}^{\cF,m+1}({\bf w})=
{\Br w}_{\{a_ia_j\}}^{{\Br\gF}_{m+1}}({\bf\Br w})$
where ${\Br\gF}_{m+1}$ is the forest on $\cD$ induced by $\gF_{m+1}$,
and ${\bf\Br w}={({\Br w}_{\Br l})}_{{\Br l}\in{\Br\gF}_{m+1}}$ is defined by
${\Br w}_{\Br l}=w_l$
where $l$ is the unique link in $\gF_{m+1}\backslash\gF_m$ inducing the link
$\Br l$ in $\cD$. The functions
${\Br w}_{\Br l}^{\Br\gF}({\bf\Br w})$ are defined by the same algorithm
as in theorem II.2 but with $\cD$ instead of $I_n$ as a point set, and
with the following {\sl induced choice of root}.
If $\Br C$ is a non empty subset or cluster of $\cD$,
${\Br C}=\{a_1,\ldots,a_\mu\}$,
let $C=\cup_{\nu=1}^\mu a_\nu$. It is a non empty subset of $I_n$ and
already has a chosen root $r_C$. Then $\exists!\ \nu,1\le\nu\le\mu$
such that $r_C\in a_\nu$.
So take ${\Br r}_{\Br C}=a_\nu\in{\Br C}$ to be the {\sl induced root}
of $\Br C$, that will get involved in the definition of the
${\Br w}_{\Br l}^{\Br\gF}({\bf\Br w})$.

Finally, and this is the last ingredient that allows us to copy the
precedent proof, we have forced $w_{\{ij\}}^{\cF,m+1}({\bf w})=1$
if $i$ and $j$ fall in the same component of $\cD$ so that we once again
recover the missing factor
$$
\exp\Biggl(\sum_{{\{ij\}\in\Pn} \atop {\{ij\}\sharp\cD}}u_{\{ij\}}^{m+1}\Biggr)
\ \ .
\eqno({\rm IV.10})
$$
\endproof
\medskip
The reader must have noticed along this proof that we used a fairly general
recipe to obtain a jungle formula from a forest formula.
We also have a recipe to get Taylor interpolation formulas from algebraic ones:
simply apply them to differentiation operators for the $u_l$'s.

Let $\Pnm=\Pn\times I_m$ and $\cS$ be the space of smooth functions from
$\RR^\Pnm$ to an arbitrary Banach space $\cV$.
An element of $\RR^\Pnm$ will be generally denoted by ${\bf x}={(x_l^k)}_
{(l,k)\in\Pnm}$. The vector with all entries equal to $1$ will be noted
$\bbbone$. $H$ is an arbitrary element of $\cS$.
We can state the following theorems, whose proofs are rewritings of that of
theorem III.1.

\theorem{IV.3 (The Brydges-Kennedy Taylor jungle formula)}{
$$
H(\bbbone)=
\sum_{{\cF=(\gF_1,\ldots,\gF_m)} \atop {m-{\rm jungle}}}
\biggl(\prod_{l\in\gF_m}\int_0^1 dh_l\biggr)
\Biggl(\biggl(\prod_{k=1}^m\Bigl(\prod_{l\in\gF_k\backslash\gF_{k-1}}
{\partial \over {\partial x_l^k}}\Bigr)\biggr)H\Biggr)
\bigl(X_\cF^{BK}({\bf h})\bigr) \ \ .
\eqno({\rm IV.11})
$$
\noindent Here $X_\cF^{BK}({\bf h})$ is the vector ${(x_l^k)}_{(l,k)\in\Pnm}$
of $\RR^\Pnm$ defined by $x_l^k=h_l^{\cF,k}({\bf h})$, which is the value
at which we evaluate the
complicated derivative of $H$.}

and

\theorem{IV.4 (The rooted Taylor jungle formula)}{
$$
H(\bbbone)=
\sum_{{\cF=(\gF_1,\ldots,\gF_m)} \atop {m-{\rm jungle}}}
\biggl(\prod_{l\in\gF_m}\int_0^1 dw_l\biggr)
\Biggl(\biggl(\prod_{k=1}^m\Bigl(\prod_{l\in\gF_k\backslash\gF_{k-1}}
{\partial \over {\partial x_l^k}}\Bigr)\biggr)H\Biggr)
\bigl(X_\cF^r({\bf w})\bigr) \ \ .
\eqno({\rm IV.12})
$$
\noindent Here $X_\cF^r({\bf w})$ is the vector ${(x_l^k)}_{(l,k)\in\Pnm}$
of $\RR^\Pnm$ defined by $x_l^k=w_l^{\cF,k}({\bf w})$, which is the value
at which we evaluate the
complicated derivative of $H$.}

Let us conclude this section by recalling the important
positivity property of the Brydges-Kennedy forest and jungle formulas.

\theorem{IV.5 (Positivity)}
{Let $\cF$ be a $m$-jungle, and
$M^{k}$, $1\le k\le m$ be any
sequence of $m$ positive symmetric $n$ by $n$ matrices with entries
$m^{k}_{ij}$. Then the interpolated
matrices $M^{k}({\bf h})$ with entries
$m^{k}_{ij}({\bf h})= h_{ij}^{\cF,k}({\bf h}) m^{k}_{ij}$ if $i<j$ and
$m^{k}_{ii}({\bf h})= m^{k}_{ii}$ are all positive.
}

\prf
The definition of $M^k({\bf h})$ only involves the forests $\gF_{k-1}$ and
$\gF_k$ and the $h_l$'s for $l$ in $\gF_k$.
Note that, for every link $l$,
$h_l^{\cF,k}({\bf h})=h_l^{\gF_k}({\Br{\bf h}})$,
the right hand side is the $h_l^{\gF_k}({\Br{\bf h}})$ function of the simple
forest formalism of Section II, where ${\Br{\bf h}}={({\Br h}_l)}_{l\in\gF_k}$
with ${\Br h}_l=h_l$ if $l\in\gF_k\backslash\gF_{k-1}$
and ${\Br h}_l=1$ if $l\in\gF_{k-1}$.
In fact, for all points $i$ and $j$ connected by $\gF_{k-1}$, i.e. such that
$L_{\gF_k}\{ij\}\subset\gF_{k-1}$, we have
$$
h_l^{\gF_k}({\Br{\bf h}})=\inf\bigl\{{\Br h}_l,\ l\in L_{\gF_k}\{ij\}\bigr\}
=\inf \{1\}
=1{{{\rm def}\atop =}\atop{ }}h_l^{\cF,k}({\bf h})\ \ .
\eqno({\rm IV.13})
$$
\indent
So we just need to prove the positivity in the simple forest formalism.
This has already been done by Brydges in [B2], but we give here a
proof along the lines of our derivation of the formula (II.1).

Suppose we have a forest $\gF$ with parameters ${\bf h}={(h_l)}_{l\in\gF}$
between 0 and 1, and a positive symmetric matrix $M$ with entries
$m_{ij}$ that is interpolated by $M({\bf h})$ with entries $m_{ij}({\bf h})
=m_{ij}h_l^\gF({\bf h})$.
Let us number the links of $\gF$ into an o-forest $F=(l_1,\ldots.l_\tau)$
in order that $1\ge h_{l_1}\ge\ldots\ge h_{l_\tau}\ge 0$ .
We perform the same change of variables as in the formula (II.23).
If $i\ne j$ let us denote by $U_{\{ij\}}$ the matrix with zero entries
except at the locations $(i,j)$ and $(j,i)$ where we put 1.
Let $M_{\rm diag}$ be the matrix with entries $m_{ii}\delta_{ij}$ the diagonal
part of $M$.
Then
$$
M({\bf h})=M_{\rm diag}+\sum_{l\in\Pn}h_l^F({\bf h}).U_l m_l
\eqno({\rm IV.14})
$$
and for the same reason as (II.23)
$$
M({\bf h})=M_{\rm diag}+\sum_{\nu=1}^\tau t_\nu M_\nu^F
\eqno({\rm IV.15})
$$
where $M_\nu^F=\sum_{\{ij\}/\nu}U_{\{ij\}}m_{ij}$, and finally
$$
M({\bf h})=(1-\sum_{\nu=1}^\tau t_\nu)M_{\rm diag}+
\sum_{\nu=1}^\tau t_\nu(M_\nu^F+M_{\rm diag})
\ \ .
\eqno({\rm IV.16})
$$

Now note that for $1\le\nu\le\tau$, $t_\nu\ge0$ and $\sum_{\nu=1}^\tau
t_\nu=h_{l_1}\le1$ so $1-\sum_{\nu=1}^\tau t_\nu\ge0$, and $M_\nu^F+
M_{\rm diag}$
is a positive symmetric matrix.
Indeed let us fix $\nu$ and consider the o-forest $F_\nu=(l_1,\ldots,l_\nu)$
completed at stage $\nu$ from the given o-forest $F$. We can reorder the base
vectors so that $M$ looks like a matrix with blocks corresponding to indexes
in the same component of $F_\nu$. Taking only the diagonal blocks and putting
the others to 0 we get exactly the matrix $M_\nu^F+M_{\rm diag}$.
It is clear that such a process preserves the positivity of the matrix.
$M({\bf h})$, a sum of positive matrices with non negative
multipliers, is positive.
\endproof

Note that the rooted formulas {\bf do not preserve positivity} as can be seen
on simple examples, with 4 points for instance.


\bigskip
\noindent{\bf V. Some concrete examples}
\medskip

\medskip
\noindent{\bf V.A) Gaussian measures perturbed by a small interaction}
\medskip
In constructive field theory cluster expansions are typically used
to perform the thermodynamic
limit of a field theory with cutoffs and
a small local interaction $I_{\La}(\ph)$
in a finite volume $\La$. The cluster expansion expresses the partition
function as a polymer gas with hard core conditions (see e.g. [R2]).
The set $I_{n}$ is then made of a partition of $\La$ into (unit size) cubes,
and clusters are subsets of such cubes.

For instance  consider the free bosonic Gaussian measure
$d\mu_{C}$ in $\RR^{d}$ defined by a covariance $C$ with ultraviolet cutoff
and good decrease at infinity.
A standard example is
$$  C(x,y) = {1\over (4\pi)^{d/2}}\int_{1}^{+\infty}
{d\al \over \al^{d/2}} e^{-\al m^{2} - |x-y|^{2}/4\al}
\eqno({\rm V.A.1})$$
so that
$$   \hat C(p) = {e^{-(p^{2}+m^{2})}\over
p^{2}+m^{2}}
\eqno({\rm V.A.2})$$

We perturb this free theory  by adding an interaction such as
$e^{-g\int_{\La}\ph^{4}(x) } dx $, and we want to
perform the thermodynamic limit
$\La \to \infty$, that is to define and study an intensive quantity
such as the pressure
$$
p = \lim_{\La \to \RR^{d}}
{1\over | \Lambda |} \log Z(\Lambda) \ , \eqno({\rm V.A.3})$$
where the partition function $ Z(\Lambda)$ in a finite volume $\Lambda$ is
$$
Z(\Lambda)= \int d\mu_{C}(\phi)e^{-g\int_\Lambda\phi^4(x)dx} \ ,
\eqno({\rm V.A.4})
$$

Let us explain how the Taylor formula (III.2) performs the task
of rewriting the partition function as a dilute gas
of clusters with hard core interaction.
We write $\La= \cup_{i\in I_{n}}b_{i}$, where each $b_{i}$ is a unit cube,
and define $\ch_{b}$ as the characteristic function of $b$, and $\ch_{\La}=
\sum_{i\in I_{n}} \ch_{b_{i}} $.
Since the interaction lies entirely within $\La$,
the covariance $C$ in (V.A.4)
can be replaced by
$C_{\La}=  \ch_{\La}(x) C(x,y) \ch_{\La}(y)$ without changing the value
of $Z(\La)$. Moreover $C_{\La}$ can be
interpolated, defining for $l=\{i,j\} \in \Pn$
$$
C_{\La}({(x_l)}_{l\in\Pn})(x,y) =  \sum_{i=1}^{n}\ch_{b_{i}}(x)
C(x,y) \ch_{b_{i}}(y)$$
$$ + \sum_{\{i,j\}\in \Pn} x_{\{ij\}}
\bigl(\ch_{b_{i}}(x)  C(x,y) \ch_{b_{j}}(y)+
\ch_{b_{j}}(x)  C(x,y) \ch_{b_{i}}(y) \bigr)
\eqno({\rm V.A.5})
$$
Remark that $C_{\La}(1,...,1) = C_{\La}$.
Now we apply the  Taylor formula (III.2) with the function
$H$ being the partition function obtained by replacing in (V.A.3-4)
the covariance $C$ by $C_{\La}({(x_l)}_{l\in\Pn})$.
Here it is crucial to use the positivity theorem IV.5, in order
for the interpolated
covariance to remain positive, hence for the
corresponding normalized Gaussian measure
to remain well defined. From the rules of Gaussian integration
of polynomials, we can compute the effect of deriving
with respect to a given $x_{l}$ parameter, and we obtain
that (III.2) in this case takes the form
$$
Z(\La)= H(\bbbone) =
\sum_{\gF}\int d\mu_{C_{\La}(X_{\gF}^{BK}({\bf h})) }(\ph )
\biggl( \prod_{l\in \gF} \int_{0}^{1} dh_{l} \biggr)
$$
$$\biggl\{\prod_{l =\{ij\} \in \gF} \int dx dy \ch_{b_{i}}(x)
\ch_{b_{j}}(y)C(x,y)
 {\de \over \de \ph (x) }{\de \over \de \ph (y) }
\biggr\} e^{-g\int_\Lambda\phi^4(x)dx}  \eqno({\rm V.A.6})
$$
Since both the local interaction
and the covariance as a matrix
factorize over the  clusters of the forest $\gF$,
the corresponding contributions in (V.A.6)
themselves factorize, which means that (V.A.6) can also be rewritten as
a gas of non-overlapping clusters, each of which has an amplitude
given by a {\it tree formula}:
$$
Z(\La)= \int d\mu_{C_{\La} }(\ph )e^{I_{\La}(\ph)}=
 \sum_{{{\rm sets\ }\{Y_{1},...,Y_{n}\}} \atop
{Y_{i} \cap Y_{j} = \emptyset, \cup Y_{i} =\La}} \prod_{i=1}^{n} A(Y_{i})
\eqno({\rm V.A.7})
$$
$$
A(Y) =
\sum_{\gT \ {\rm on }\ Y} \bigl(
\prod_{l\in \gT}\int_{0}^{1} dh_{l} \bigr)
\int d\mu_{C_{Y}(X_{\gT}^{BK}({\bf h})) }(\ph )
$$
$$\biggl\{\prod_{l =\{ij\} \in \gT} \int dx dy \ch_{b_{i}}(x)
\ch_{b_{j}}(y)C(x,y)
 {\de \over \de \ph (x) }{\de \over \de \ph (y) }
\biggr\} e^{-g\int_Y\phi^4(x)dx}
\eqno({\rm V.A.8})
$$
where $b_{i}$ and $b_{j}$ are the two ends of the line $l$, and the sum
is over trees $\gT$ which connect together the set $Y$, hence have exactly
$|Y|-1$ elements (if $|Y|=1$, $\gT=\emptyset$ connects $Y$).
The measure $ d\mu_{C_{Y}(\{X_{\gT}^{BK}({\bf h})\}) }(\ph)$
is the normalized Gaussian measure
with (positive) covariance
$$
C_{Y}(X_{\gT}^{BK}({\bf h}))(x,y) = \ch_{Y}(x)\bigl(h_{\gT}({\bf h})(x,y)\bigr)
C(x,y) \ch_{Y}(y)
\eqno({\rm V.A.9})
$$
where $h_{\gT}({\bf h})(x,y)$ is 1 if $x$ and $y$ belong to the same cube,
and otherwise it
is the infimum of the parameters $h_{l}$ for $l$ in the unique path
$L_{\gT}(b(x),b(y))$ which in the tree $\gT$
joins the cube $b(x)$ containing $x$ to the cube
$b(y)$ containing $y$.

Formula (V.A.8) is somewhat shorter than the different
formulas of [R1-2], and can be used in the same way
to check that given any constant $K$, for
small enough $g$ with ${\rm Re} \ g >0$
$$
\sum_{Y \ {\rm st}\ 0\in Y} |A(Y)| K^{|Y|} \le 1
\eqno({\rm V.A.10})
$$
Proof of (V.A.10) requires the slightly cumbersome computation of the
action of the functional derivatives in (V.A.8) and a bound on the resulting
functional integral. The method is identical to [R1-2]. Remark that although
the full amplitudes $A(Y)$ defined in (V.A.8) must be identical
to those in [R1-2], the subcontributions associated to particular trees
are different.

The Mayer expansion defined below allows to deduce from (V.A.10)
the existence and e.g. the Borel summability
in $g$ of thermodynamic functions such as the pressure $p$
defined by (V.A.3).

\bigskip
\noindent{\bf B) The Mayer expansion}
\medskip

In the cluster expansion (V.A.7), the condition
that the disjoint union of all clusters
is $\La$ is a global annoying constraint.
Remark that the polymer
amplitudes are translation invariant. In particular
the trivial amplitude of a singleton cluster $Y=\{b\}$ is a number $A_{0}$
independent of $b$. Redefining $A_{r}(Y)= A(Y)/A_{0}^{|Y|}$
and $Z_{r}(\La) = Z(\La)/A_{0}^{|\La|}$ we quotient
out all the trivial clusters so that
$$
Z_{r}(\La)= 1+
\sum_{n\ge 1} \sum_{{\rm sets }\ \{Y_{1},...,Y_{n}\} \atop
|Y_{i}| \ge 2 \ , \ Y_{i} \cap Y_{j} = \emptyset}
\prod_{i=1}^{n} A_{r}(Y_{i})
\eqno({\rm V.B.1})
$$
This is the partition function
of a polymer gas: the sums over individual polymers would be
independent were it
not for the hard core constraints $ Y_{i} \cap Y_{j} = \emptyset$.
Adding in an infinite number of vanishing terms,
we can replace the sum in (V.B.1)
by a sum over ordered sequences $(Y_{1},...,Y_{n})$ of polymers
with hard core interaction and
a symmetrizing factor $1/n!$ coming from the replacement of sets by sequences.
$$
Z_{r}(\La)= 1+
\sum_{n\ge 1}  {1\over n!}\sum_{ {\rm sequences }\ (Y_{1},...,Y_{n}) \atop
|Y_{i}| \ge 2 \ }
\prod_{i=1}^{n} A_{r}(Y_{i}) \prod_{1\le i< j\le n} e^{-V(Y_{i}, Y_{j})}
\eqno({\rm V.B.2})
$$
where the hard core interaction is $V(X,Y)=0$ if $X \cap Y
=\emptyset$, and $V(X,Y)=+\infty$ if $X \cap Y
\not =\emptyset$
To factorize again this formula
we cannot apply directly the algebraic Brydges-Kennedy
formula (II.1), because the interaction $V$ can be infinite,
but the Taylor forest
formula (III.2) easily does the job.
More precisely we
define now $I_{n}$ as our set of indices, and
define $\ep^Y_{\{ij\}} = (e^{-V(X_{i}, X_{j})}-1)$, for $i \ne j$.
For a fixed sequence $(Y_1,\ldots,Y_n)$ of polymers, consider the function
$$
H({(x_l)}_{l\in\Pn}) = \prod_{l \in \Pn} (1 + x_l\ep_l^Y)
\eqno({\rm V.B.3})
$$
Rewrite (V.B.2) as
$$
Z_{r}(\La)= 1+
\sum_{n\ge 1}  {1\over n!}\sum_{ {\rm sequences }\ (Y_{1},...,Y_{n}) \atop
|Y_{i}| \ge 2 \ }
 H(\bbbone ) \prod_{i=1}^{n} A_{r}(Y_{i})
$$
$$ = 1 + \sum_{n\ge 1}  {1 \over n!}
\sum_{ {\rm sequences }\ (Y_{1},...,Y_{n}) \atop
|Y_{i}| \ge 2 \ } \prod_{i=1}^{n} A_{r}(Y_{i}) \sum_{\gF}
\biggl( \prod_{l\in \gF} \int_{0}^{1} dh_{l} \biggr)
\biggl( \prod_{l\in \gF} \ep^Y_{l} \biggr)
\biggl( \prod_{l\not\in \gF} (1+ h_{l}^{\gF} ({\bf h}) \ep^Y_{l}) \biggr)
$$
$$=  \sum_{n\ge 0}  {1 \over n!}\biggl(
\sum_{k\ge 1}  {1 \over k!}\sum_{ {\rm sequences }\ (Y_{1},...,Y_{k}) \atop
|Y_{i}| \ge 2 \ } \Bigl(\prod_{i=1}^{k} A_{r}(Y_{i})\Bigr)
C^{T}(Y_{1},...,Y_{k})
\biggr)^{n}
\eqno({\rm V.B.4})
$$
where
$$
C^{T}(Y_{1},...,Y_{k}) =
\sum_{{G \ {\rm connected\ graph}}\atop{{\rm on} \{1,....,k\}}}
\prod_{l\in G} \ep^Y_{l}
$$
$$ = \sum_{\gT \ {\rm tree\ on \ }  \{1,....,k\}}
\biggl( \prod_{l\in \gT} \int_{0}^{1 } dh_{l}\biggr)
\biggl(\prod_{l\in {\cal P}_{k}, \ l\in \gT}  \ep^Y_{l}\biggr)
\prod_{l\in {\cal P}_{k}, \ l \not \in \gT} \bigl(1 + h_{l}^{\gT}({\bf h})
\ep_{l}^Y \bigr)
\eqno({\rm V.B.5})
$$
where $ h_{l}^{\gT}({\bf h}) $ is, if $l=\{ij\}$, the
infimum of the parameters $h_{l'}$ for $l'$ in the unique path
$L_{\gT}\{ij\}$ which in the tree $\gT$
joins $i $ to $j$.

We obtain immediately that
$$
\log Z_{r}(\La) =
\sum_{k\ge 1}  {1 \over k!}\sum_{ {\rm sequences }\ (Y_{1},...,Y_{k}) \atop
|Y_{i}| \ge 2 \ } \biggl(\prod_{i=1}^{k} A_{r}(Y_{i})\biggr)
C^{T}(Y_{1},...,Y_{k})
\eqno({\rm V.B.6})
$$
Formulas (V.B.5-6) are more explicit than
those used in [R1-2]
and have all desired
advantages (every tree coefficient forces the necessary links
and is bounded by 1, since $| (1 + h_{l}^{\gT}({\bf h}) \ep^Y_{l})|\le 1$).
They can therefore be used together with (V.A.10)
to control in a similar way the thermodynamic
function
$p = A_{0} + \lim _{\La \to \RR^{d}} {1 \over |\La |}
\log  Z_{r}(\La)$.

\bigskip
\noindent
{\bf C) A single formula for the succession of a cluster and a Mayer expansion}
\medskip

Formula (V.B.6) involves trees and forests on two different kinds of objects.
First are the links between boxes in our lattice that make the clusters $Y_i$.
Then we have Mayer links between indexes $i$ in $I_k$.
This leads to complications, if we iterate the process, when doing a
renormalization group analysis. So we propose a formula with two types of
links but on the same objects, it is a $2$-jungle formula.
For simplicity we write it  in the algebraic setting of section II, i.e.
with $Z=\exp(\sum u_l)$ instead of $Z_r(\La)$, and perform the apparently
stupid operation $\log\exp(\sum u_l)$ !

The links $l$ are between elements of $I_n$, but now we introduce a new set of
objects $\cD_n=I_n\times\NN$. If ${\Br l}$ is a link $\{(b,k),(b',k')\}$
between different elements of $\cD_n$ we let $u_{\Br l}=u_{\{bb'\}}$
and $\delta_{\Br l}=\delta_{bb'}$ where the second delta is Kroneker's.
Let $\cF=(\gF_1,\gF_2)$ be a $2$-jungle on $\cD_n$, we say that it is
{\sl nice} if it fulfills the following requirements:

- $\gF_1$ is only made of horizontal links i.e. of the form $\{(b,k),(b',k)\}$;

- $\gF_2\backslash\gF_1$ is only made of vertical links of length $1$
i.e. of the form $\{(b,k),(b,k+1)\}$;

- $\gF_2$ is a non empty tree;

- there are no singletons among the clusters of $\gF_1$ in the support of
$\gF_2$;

- there is a unique cluster of $\gF_1$ in the support of $\gF_2$ that lies in
the $0$-th level $I_n\times\{0\}$;

- if we fix a root $r$ of $\gF_2$ in that unique cluster, then for every
element $(b,k)$ in the support of $\gF_2$, $k$ is the number of vertical
links on the path going from $(b,k)$ to $r$ (this property is independent of
the choice of $r$).

We denote by $k(\cF)$ the number of clusters of $\gF_1$ in the support of
$\gF_2$. We introduce the following lexicographical order on $\cD_n$,
$(b,k)\preceq(b',k')$ if and only if
$k<k'$ or ($k=k'$ and $b\le b'$). This is the order we use to choose a root
for each non empty finite subset of $\cD_n$ according to the rooted jungle
formalism.
We let ${\bf h}={(h_{\Br l})}_{{\Br l}\in\gF_1}$ and
${\bf w}={(w_{\Br l})}_{{\Br l}\in\gF_2\backslash\gF_1}$ and we take the
compound $({\bf h},{\bf w})$ instead of the $\bf h$ or the $\bf w$
vectors that were used in (IV.1) and (IV.9). The functions $h_{\Br l}^{\cF,1}$
and $w_{\Br l}^{\cF,2}$ are however defined exactly in the same manner as in
section IV. We now claim that
\theorem{V.C.1}{
$$
\log Z=
\sum_{{\cF=(\gF_1,\gF_2)}\atop{\rm nice\ 2-jungle}}{1\over{k(\cF)}}
\biggl(\prod_{{\Br l}\in\gF_1}u_{\Br l}\int_0^1dh_{\Br l}\biggr)
\biggl(\prod_{{\Br l}\in\gF_2\backslash\gF_1}
(-\delta_{\Br l})\int_0^1dw_{\Br l}\biggr)
$$
$$
\times
\exp\biggl(\sum_{\Br l}h_{\Br l}^{\cF,1}({\bf h},{\bf w})u_{\Br l}\biggr)
\ .\
\prod_{{\Br l}\notin\gF_2\backslash\gF_1}
\Bigl(1-w_{\Br l}^{\cF,2}({\bf h},{\bf w})\delta_{\Br l}\Bigr)
\ \ \ .
\eqno({\rm V.C.1})
$$
}
\prf

For $m\in\NN$, let $\cD_{n,m}=I_n\times\{0,1,\ldots,m\}\subset\cD_n$.
As in (V.B.6) we have
$$
\log Z=\lim_{m\to+\infty}L_m
\eqno({\rm V.C.2})
$$
where
$$
L_m=\sum_{s=1}^m{1\over{s!}}\sum_{{{\rm sequences\ }(Y_1,\ldots,Y_s)}
\atop {Y_i\subset I_n,\ |Y_i|\ge2}}
\biggl(\prod_{i=1}^sA(Y_i)\biggr)C^T(Y_1,\ldots,Y_s)
\eqno({\rm V.C.3})
$$
and
$$
A(Y)=\sum_{\gT\ {\rm tree\ on\ }Y}
\biggl(\prod_{l\in\gT}u_l\int_0^1dh_l\biggr)
\exp\biggl(\sum_{l\ {\rm in\ }Y}h_l^\gT({\bf h})u_l\biggr)
\ \ .
\eqno({\rm V.C.4})
$$
\indent
This time we compute $C^T(Y_1,\ldots,Y_s)$ thanks to the rooted forest
formalism. Note that
$$
\prod_{\{ij\}\subset I_n}(1+\ep_{\{ij\}}^Y)=
\prod_{\{ij\}\subset I_n}\prod_{(b,b')\in Y_i\times Y_j}(1-x_{\{ij\}}
\delta_{bb'})
\eqno({\rm V.C.5})
$$
with all the  $x_{\{ij\}}$ set to 1.
Then we apply the rooted Taylor forest formula III.3 and collect the
connected parts. We get
$$
C^T(Y_1,\ldots,Y_S)=\sum_{\gT\ {\rm tree\ on\ }I_s}
\cC(s;Y_1,\ldots,Y_s;\gT)
\eqno({\rm V.C.6})
$$
where
$$
\cC(s;Y_1,\ldots,Y_s;\gT)=\biggl(\prod_{l\in\gT}\int_0^1dw_l\biggr)
\ .
$$
$$
\Biggl(\prod_{\{ij\}\in\gT}\biggl(\sum_{(b,b')\in Y_i\times Y_j}
(-\delta_{bb'})\prod_{{(c,c')\in Y_i\times Y_j}\atop{(c,c')\ne(b,b')}}
(1-w_{\{ij\}}^\gT({\bf w})\delta_{cc'})\biggr)\Biggr)
$$
$$
.\ \prod_{\{ij\}\notin\gT}
\biggl(\prod_{(b,b')\in Y_i\times Y_j}
(1-w_{\{ij\}}^\gT({\bf w})\delta_{bb'})\biggr)
\ \ .
\eqno({\rm V.C.7})
$$
\indent
The only trees $\gT$ giving a non zero contribution are those where for
any $i\ne j$ in the same layer of $\gT$, $Y_i\cap Y_j=\emptyset$.
In fact for such $i$ and $j$, $\{ij\}\notin\gT$ and
$w_{\{ij\}}^\gT({\bf w})=1$ so in (V.C.7) we get a factor
$$
\prod_{(b,b')\in Y_i\times Y_j}(1-\delta_{bb'})
=e^{-V(Y_i,Y_j)}
\eqno({\rm V.C.8})
$$
forcing the non overlapping condition.

Given a sequence $(Y_1,\ldots,Y_s)$ and a tree $\gT$ satisfying that
property, we can define for $i\in I_k$,
${\Br Y}_i=Y_i\times\{l^\gT(i)\}\subset\cD_{n,m}$.
Here $l^\gT(i)$ is the height of vertex $i$ in the tree $\gT$.
The ${\Br Y}_i$ are all disjoint.

For a finite subset $\Br Y$ of $\cD_n$ let
$$
{\Br A}({\Br Y})=
\sum_{\gT\ {\rm tree\ on\ }{\Br Y}}
\biggl(\prod_{{\Br l}\in\gT}u_{\Br l}\int_0^1dh_{\Br l}\biggr)
\exp(\sum_{{\Br l}\in{\Br Y}}h_{\Br l}^\gT({\bf h})u_{\Br l}\biggr)
\eqno({\rm V.C.9})
$$
so that $A(Y_i)={\Br A}({\Br Y}_i)$ for every $i\in I_s$.
Now define the (unordered) set $\cO=\cO(Y_1,\ldots,Y_s;\gT)=
\{{\Br Y}_1,\ldots,{\Br Y}_s\}$.
It is easy to see that $\cO$ is a non empty set of disjoint polymers
($|{\Br Y}|\ge2$ for each ${\Br Y}\in\cO$)
lying in $\cD_{n,m}$, with exactly one element in the ground level
$I_n\times\{0\}$, furthermore, the labels $k$ of the occupied levels
$I_n\times\{k\}$ form an interval $\{0,1,\ldots,q\}$, $q\le|\cO|-1$.
An $\cO$ verifying these properties is called {\sl admissible}.

We will sum over admissible $\cO$'s in (V.C.3)
$$
L_m=\sum_{{\rm admissible\ }\cO}{1\over{|\cO|!}}
\biggl(\prod_{{\Br Y}\in\cO}{\Br A}({\Br Y})\biggr)
\sum_{{(Y_1,\ldots,Y_s);\gT}\atop{{\rm st\ }\cO=\cO(Y_1,\ldots,Y_s;\gT)}}
\cC(s;Y_1,\ldots,Y_s;\gT)\ \ .
\eqno({\rm V.C.10})
$$
\indent
Knowing the sequence $Y_1,\ldots,Y_s)$, $\gT$ naturally induces a tree
on $\cO$, ${\Br\gT}={\Br\gT}(Y_1,\ldots,Y_s;\gT)$ by the rule
$\{ij\}\in\gT$ if and only if $\{{\Br Y}_i,{\Br Y}_j\}\in{\Br\gT}$.
The tree $\Br\gT$ has a natural root i.e. the unique $\Br Y$ of $\cO$
lying in the 0-th level, moreover, for each $\Br Y$ the label of the
level it belongs to is just its height in the tree $\Br\gT$ with the
mentioned choice of root. Such a tree will also be called {\sl admissible}.

Let for any tree $\Br\gT$ on $\cO$
$$
U(\cO,{\Br\gT})=\prod_{L\in{\Br\gT}}\int_0^1dw_L
\Biggl(\prod_{\{{\Br Y}\ {\Br Y}'\}\in{\Br\gT}}
\biggl(\sum_{({\Br b},{\Br b}')\in {\Br Y}\times {\Br Y}'}
(-\delta_{\{{\Br b}\ {\Br b}'\}})
\prod_{{({\Br c},{\Br c}')\in {\Br Y}\times {\Br Y}'}
\atop{({\Br c},{\Br c}')\ne({\Br b},{\Br b}')}}
(1-w_{\{{\Br Y}\ {\Br Y}'\}}^{\Br\gT}({\bf w})
\delta_{\{{\Br c}\ {\Br c}'\}})\biggr)\Biggr)
$$
$$
.\ \prod_{{\{{\Br Y}\ {\Br Y}'\}\subset\cO}\atop
{\{{\Br Y}\ {\Br Y}'\}\notin{\Br\gT}}}
\prod_{({\Br b},{\Br b}')\in {\Br Y}\times {\Br Y}'}
(1-w_{\{{\Br Y}\ {\Br Y}'\}}^{\Br\gT}({\bf w})\delta_{\{{\Br b}\ {\Br b}'\}})
\ \ .
\eqno({\rm V.C.11})
$$
and
$$
N(\cO,{\Br\gT})=
\sum_{{(Y_1,\ldots,Y_s);\gT}\atop
{{{\rm st\ }\cO=\cO(Y_1,\ldots,Y_s;\gT)}
\atop{{\rm and\ }{\Br\gT}={\Br\gT}(Y_1,\ldots,Y_s;\gT)}}}
1\ \ \ .
\eqno({\rm V.C.12})
$$
\indent
It is clear now that
$$
L_m=\sum_{{\rm admissible\ }\cO}{1\over{|\cO|!}}
\biggl(\prod_{{\Br Y}\in\cO}{\Br A}({\Br Y})\biggr)
\sum_{{\Br\gT}{\rm tree\ on\ }\cO}U(\cO,{\Br\gT}).N(\cO,{\Br\gT})
\ \ .
\eqno({\rm V.C.13})
$$
Let us admit for the moment the following
\lemma{V.C.1}{
The combinatoric factor $N(\cO,{\Br\gT})$ is $(|\cO|-1)!$ for
any admissible $\cO$ and $\Br\gT$.
}
Then
$$
L_m=\sum_{{{\rm admissible\ }\cO}\atop{{\Br\gT}{\rm\ tree\ on\ \cO}}}
{1\over{|\cO|}}
\biggl(\prod_{{\Br Y}\in\cO}{\Br A}({\Br Y})\biggr)
\sum_{{\Br\gT}{\rm\ tree\ on\ }\cO}U(\cO,{\Br\gT})
\ \ .
\eqno({\rm V.C.14})
$$
\indent
Besides,
$$
\prod_{{\Br Y}\in\cO}{\Br A}({\Br Y})=
\sum_{\gF_1}\biggl(\prod_{{\Br l}\in\gF_1}u_{\Br l}\int_0^1dh_{\Br l}\biggr)
\exp\biggl(\sum_{\Br l}h_{\Br l}^{\gF_1}({\bf h})u_{\Br l}\biggr)
\ \ ,
\eqno({\rm V.C.15})
$$
where we sum over all forests $\gF_1$ made of horizontal links $u_{\Br l}$
such that the connected components that are not singletons are exactly
the elements of $\cO$.

Finally remark that summing over trees $\Br\gT$ on $\cO$ and, for every
$\{{\Br Y},{\Br Y}'\}\in{\Br\gT}$, over $({\Br b},{\Br b}')\in
{\Br Y}\times{\Br Y}'$ is the same as summing over forests
$\gF_2\backslash\gF_1$ of vertical links $-\delta_{\Br l}$
connecting in a tree the clusters ${\Br Y}\in\cO$ formed by $\gF_1$.
It is now a matter of (rather unpleasant) checking the definitions
to convince oneself that (V.C.14) is just the result of formula (V.C.1),
except that the clusters are confined to $\cD_{n,m}$. However taking
the limit $m\to+\infty$ removes this restriction.
\endproof
\medskip
\noindent{\bf Proof of lemma V.C.1 :}
\medskip
Since $\cO$ and $\Br\gT$ are admissible, there exists a sequence
$Y_1,\ldots,Y_s$ and a tree $\gT$ such that $\cO=\cO(Y_1,\ldots,Y_s;\gT)$
and ${\Br\gT}={\Br\gT}(Y_1,\ldots,Y_s;\gT)$.
In fact suppose we are given a one-to-one map $\sigma:\cO\mapsto I_s$
such that $\sigma({\Br Y}_1)=1$, where ${\Br Y}_1$ is the unique element
of $\cO$ lying in the ground level of $\cD_n$. We can construct a sequence
$(Y_1^\sigma,\ldots,Y_s^\sigma)$ and a tree $\gT^\sigma$ fulfilling
the previous requirements.
If $\pi$ denotes the projection of $\cD_n$ on $I_n$, we let
$Y_{\sigma({\Br Y})}^\sigma=\pi({\Br Y})$ for all ${\Br Y}\in\cO$,
 and $\gT^\sigma$ be the tree on $I_s$ induced by $\Br\gT$ through
the correspondence $\sigma$.

It is easy to see that every pair $\bigl((Y_1,\ldots,Y_s);\gT\bigr)$
we are counting is obtained that way. The interesting thing is that
there is a unique $\sigma$ giving this pair. If $\sigma$ and $\tau$
are two maps such that $\bigl(({Y_1}^\sigma,\ldots,{Y_s}^\sigma);
\gT^\sigma\bigr)=\bigl(({Y_1}^\tau,\ldots,{Y_s}^\tau);\gT^\tau\bigr)$,
$\sigma\circ\tau^{-1}$ must preserve the tree $\gT^\sigma=\gT^\tau$
, but $\sigma\circ\tau^{-1}(1)=1$ so it preserves also the root,
as a consequence it leaves the layers invariant. But, the corresponding
${\Br Y}_i$ are disjoint contained in the same level of $\cD_n$,
thus the projections $Y_i$ are distinct and $Y_{\sigma\circ\tau^{-1}(i)}
=Y_i$ forces $\sigma\circ\tau^{-1}(i)=i$ inside each layer.
In conclusion $\sigma\circ\tau^{-1}=Id$. $N(\cO,{\Br\gT})$ is the
number of $\sigma$'s that is $(s-1)!$.
\endproof

If after all that the reader is somewhat sceptical about
formula (V.C.1), he might
find it amusing to check it for $n=2$.

Now to use formula (V.C.1) for more clever applications than computing
the trivial expression
$\log\exp(\sum u_l)$, we state the analogous result for $\log Z(\La)$
in $\phi^4$ theory.
We introduce a copy of the field $\phi_k$ for each $k\in\NN$, and
write $\Phi$ for the system ${(\phi_k)}_{k\in\NN}$.
Then we define the interaction in $S(\gF_2)$, the support of $\gF_2$,
as
$$
e^{I(S(\gF_2))}(\Phi)=\prod_{(b,k)\in S(\gF_2)}
e^{-\la\int_b\phi_k^4(x)dx}
\eqno({\rm V.C.16})
$$
where we have identified the boxes $b$ in $\La$ with their labels in $I_n$.
To a link ${\Br l}=\{(b,k),(b',k')\}$ we associate the operator
$$
C_{\Br l}=\delta_{kk'}\int_{b}dx\int_{b'}dy\ C(x,y)
{{\delta}\over{\delta\phi_k(x)}}
{{\delta}\over{\delta\phi_{k'}(x)}}
\eqno({\rm V.C.17})
$$
that will play the role of $u_{\Br l}$.
$\delta_{\Br l}$ is defined in the same fashion as before.
Given a 2-jungle $\cF=(\gF_1,\gF_2)$ denote by $S_k(\gF_2)$ the $k$-th
slice $S(\gF_2)\cap(I_n\times\{k\})$ of $S(\gF_2)$.
We define the following Gaussian measure on the fields
${\phi_k}_{|S_k(\gF_2)}$ such that $k$ is an occupied level of $\gF_2$
(i.e. $S_k(\gF_2)\ne\emptyset$)
$$
d\mu_C^{\cF,{\bf h},{\bf w}}(\Phi_{|S(\gF_2)})=
\prod_{{\rm occupied\ }k}
d\mu_C^{\cF,{\bf h},{\bf w},k}({\phi_k}_{|S_k(\gF_2)})
\eqno({\rm V.C.18})
$$
where $d\mu_C^{\cF,{\bf h},{\bf w},k}({\phi_k}_{|S_k(\gF_2)})$
has covariance
$C^{\cF,{\bf h},{\bf w},k}(x,y)=C(x,y)$ if $x$ and $y$ fall in the same
$b$ with $(b,k)\in S_k(\gF_2)$; and $C^{\cF,{\bf h},{\bf w},k}(x,y)=
C(x,y).h_{\{(b,k),(b',k')\}}^{\cF,1}({\bf h},{\bf w})$ if
$x\in b$, $y\in b'$, $b\ne b'$ $(b,k)\in S_k(\gF_2)$ and
$(b',k)\in S_k(\gF_2)$.

We can now state a formula which performs
a cluster and a Mayer expansion in a single move (recall that $|\La|=n$)
\theorem{V.C.2}{
$$
\log Z(\La)=
n\log A_0+
\sum_{{\cF=(\gF_1,\gF_2)}\atop{\rm nice\ 2-jungle}}
{{A_0^{-|S(\gF_2)|}}\over{k(\cF)}}
\biggl(\prod_{{\Br l}\in\gF_1}\int_0^1dh_{\Br l}\biggr)
\biggl(\prod_{{\Br l}\in\gF_2\backslash\gF_1}
(-\delta_{\Br l})\int_0^1dw_{\Br l}\biggr)
$$
$$
.\ \prod_{{\Br l}\notin\gF_2\backslash\gF_1}
(1-w_{\Br l}^{\cF,2}({\bf h},{\bf w})\delta_{\Br l})
\int d\mu_C^{\cF,{\bf h},{\bf w}}(\Phi_{|S(\gF_2)})
\biggl(\prod_{{\Br l}\in\gF_1}C_{\Br l}\biggr)
e^{I(S(\gF_2))}\ \ .
\eqno({\rm V.C.19})
$$
}
\noindent{\bf Proof} Use first $\log Z(\La)=
n\log A_0+ \log Z_{r}(\La) $, then repeat exactly the same
line of arguments than for  Theorem V.C.1. \endproof

\noindent From Theorem V.C.2 one can construct a series for the pressure
$\lim_{\La \to \infty} |\La|^{-1}\log Z(\La) $ which is the sum of the trivial
term $\log A_0$ plus a sum over nice 2 jungles that extend
horizontally over the infinite lattice of all cubes covering $\RR^{d}$,
and such that the well defined unique root
of $\gF_{2}$ is the particular cube containing the origin.
(This requires to neglect boundary terms of order $|\La|^{(d-1)/d}$
in $\log Z(\La)$). For small
$\la$ this series is absolutely convergent.

\bigskip
\noindent{\bf D) The resolvent expansion}
\medskip
In many situations physical situations (polymers, disordered systems)
we want to compute not a full theory but a single Green's function
which is expressed as the inverse of some operator.
This mathematical situation is formally equivalent to
0-component field theory (or to supersymmetric theories)
in which the usual normalizing fermionic or bosonic
determinants have been cancelled out.

Consider a finite dimensional operator $M$ on $\RR^{n}$ with matrix elements
$m_{ij}$, and a norm strictly smaller than 1. The operator
$A= {1 \over 1+M}$ is well defined through its power series expansion. Define
$M({\bf x})$ as the matrix with elements
$x_{\{ij\}}m_{ij} $ if $i\ne j$ and diagonal elements $m_{ii}$ equal to
those of $M$, and $M_{\{ij\}}$ as the matrix with 0 elements except
${(M_{\{ij\}})}_{ij}=m_{ij}$ and ${(M_{\{ij\}})}_{ji}=m_{ji}$.
The Taylor forest formula (III.2) applied to the operator
$H({\bf x})=  {1 \over 1+M({\bf x})}$ gives
$$
A={1 \over 1+  M}= H({\bf 1}) = \sum_{\gF}
\biggl( \prod_{l\in \gF} \int_{0}^{1} dh_{l} \biggr)
\biggl( \prod_{l\in \gF}   \oint {1\over 2\pi i}
{d x_{l} \over x_{l}^{2}} \biggr)
{1 \over 1+  M(X_{\gF}({\bf h})) +
\sum_{l \in \gF } x_{l}M_{l} }
\eqno({\rm V.D.1})
$$
where the action of multiple derivatives is for compactness
rewritten as a multiple Cauchy integral, and the analyticity radii
in (V.D.1) are small enough. This does not look like a very clever rewriting,
but it allows to reblock the power series for ${1 \over 1+  M}$
according to the set of sites truly visited in this power series.
In particular if we impose a given entry to the matrix $A$ the forest
formula (V.D.1) reduces to a tree formula:
$$
A_{ij}=\biggl({1 \over 1+  M}\biggr)_{ij}= \sum_{Y \subset \{1,...,n\}
\atop {i \in Y,\ j\in Y}}
\sum_{\gT\ {\rm tree \ on }\ Y}
\biggl( \prod_{l\in\gT} \int_{0}^{1} dh_{l} \biggr)
$$
$$\biggl( \prod_{l\in\gT}   \oint {1\over 2\pi i}
{d x_{l} \over x_{l}^{2}} \biggr)
{1 \over 1_{Y}+  M_{Y}(X_{\gT}({\bf h})) +
\sum_{l \in\gT} x_{l}M_{Y,l} }
\eqno({\rm V.D.2})
$$
where $ 1_{Y}$ is the identity on $R^{Y}$,
$M_{Y}(X_{\gT}({\bf h}))$ is the matrix on $R^{Y}$ with diagonal
entries $m_{ii}$ (for $i\in Y$)
and off diagonal entries $h^{\gT}_{ij}({\bf h}) m_{ij}$
(for $i \in Y$, $j\in Y)$, and $M_{Y,l}$ is the restriction of
$M_l$ to indexes in $Y$.

This kind of formulas, eventually in combination with a large field/small
field analysis should be useful for
interacting random walks [IM] and presumably for the study
of disordered systems.

\bigskip
\noindent{\bf E) Cluster expansions with
large/small field conditions: the $m=2$ jungle.}
\medskip
Jungle formulas with several levels are interesting
when there are various type of links with  priority rules between them.
The simplest example is a single scale cluster expansion but with
so-called small/large field conditions [R2], in which one does $not$
want to derive links between cubes of a large field connected component.
Indeed large field regions are small for probabilistic reasons, but the
vertices of perturbation theory
(created by the action of functional derivatives
such as those in (V.A.8)) may not be small there, hence it would be
dangerous to blindly expand the clusters in the usual way.

We want to study a theory with partition function such as
$$
Z(\Lambda)= \sum_{\Ga \subset \La} Z(\Lambda, \Ga)\ ,
\ \ \  Z(\Lambda, \Ga) = \int d\mu_{C}(\phi)
\ch_{\Ga}(\ph) e ^{-g\int_\Lambda I(\ph) dx} \ ,
\eqno({\rm V.E.1})
$$
where $C$ is a propagator with decay at the unit scale,
and $I(\ph)$ is some local interaction.
Without describing a precise model
we shall assume that the large field condition $\ch_{\Ga}(\ph)$
is such that the functional integral $\int d\mu_{C}(\phi)
\ch_{\Ga}(\ph) e ^{-g\int_\Lambda I(\ph) dx}$ can be bounded
by $K^{|\La-\Ga|} c^{|\Ga|}$ where $K$ is fixed and $c$ is a small constant
that can tend to zero with some coupling constant
in $I(\ph)$; one also assumes that the outcome of a functional derivative
$\int_{b}dx {\de \over \de \ph(x)}$ acting on
$\ch_{\Ga}(\ph) e ^{-g\int_\Lambda I(\ph) dx}$ gives a small factor but only
in the ``small field region'', hence if  $b \not \in \Ga$.

The set $I_{n}$ is again the set of the cubes
which pave our finite volume $\La$.
We fix a given subset $\Ga \subset I_{n}$ and define, for $l=\{ij\} \in
{\cal P}_{n}$,
$\ep^{\Ga}_{l}=1$ if $b_{i}\in \Ga$, $b_{j}\in \Ga$ and
${\rm dist (b_{i}, b_{j}} \le M$. $M$ is some constant that will be fixed
to a large value.
 Otherwise we put $\ep^{\Ga}_{l}=0$.
We define also
$\et^{\Ga}_{l} =1 -\ep^{\Ga}_{ij}$. Hence
$\ep^{\Ga}_{l}=1$ means that $b_{i} $ and $b_{j}$ are ``large field cubes''
which are closer than distance
$M$, and $\et^{\Ga}_{l}=1$ means the contrary.

The level two jungle formula will at the first level create connections
whose clusters are automatically the ``connected components'' of $\Ga$
in the generalized sense that up to distance $M$ two cubes are considered
connected.
Then the second level will create ordinary connections with the propagator
$C$.
Therefore we introduce a first set of interpolation parameters
$\{x^{1}_{l}\}$ and define
$$H^{\Ga}(\{x^{1}_{l}\}) =
\prod_{l \in \Pn} (x^{1}_{l}\ep^{\Ga}_{l}+\et^{\Ga}_{l}) \ .
\eqno({\rm V.E.2})
$$
Remark that $H^{\Ga}(1,...,1)=1$, so that we can freely multiply
$Z(\La, \Ga ) $ by $H^{\Ga}(1,...,1)$ without changing its value.
 We introduce the second set
of parameters $\{x^{2}_{l}\}$ on the covariance $C$ exactly as in
(V.A.5).

Then the formula (IV.11) with $m=2$ simply gives:

$$
Z(\La, \Ga) =
\sum_{\cF= (\gF_{1}, \gF_{2})
\atop {\rm 2-jungle}} \biggl( \prod_{l \in \cF} \int_{0}^{1} dh_{l} \biggr)
\biggl( \prod_{l \in \gF_{1}} \ep^{\Ga}_{l} \biggr)
\biggl(\prod _{l \not\in \gF_{1}}
\bigl(\et^{\Ga}_{l} + \ep^{\Ga}_{l} h^{\cF,1}_{l} ({\bf h})\bigr)
\biggr)
$$
$$
\int d\mu_{C_{\cF}({\bf h})} (\ph)\prod_{l=\{ij\}
\in \gF_{2}-\gF_{1}} \biggl\{ \int dx dy \ch_{b_{i}}(x)
\ch_{b_{j}}(y)C(x,y)
 {\de \over \de \ph (x) }{\de \over \de \ph (y) }
\biggr\} \ch_{\Ga}(\ph) e ^{-g\int_\Lambda I(\ph) dx}
\eqno({\rm V.E.3})
$$
where $ d\mu_{C_{\cF}({\bf h})}$ is the normalized Gaussian measure with
positive covariance
$$
C_{\cF}({\bf h})(x,y) = h_{l(x,y)}^{\cF,2}({\bf h})
C(x,y)
\eqno({\rm V.E.4})
$$
where $ h_{l(x,y)}^{\cF,2}({\bf h})$ is by definition 1 if $x$ and
$y$ belong to the same cube, and is
$ h_{l}^{\cF,2}({\bf h})$ if $l=\{ij\}$, $x\in b_{i}$
and $y\in b_{j}$ or $x\in b_{j}$
and $y\in b_{i}$.

The only non-zero terms in this formula are those for which the
clusters associated to the forest $\gF_{1}$ are exactly the set
of ``connected components'' $\Ga_{a}$ of the large field region
in the generalized sense (for $M=0$ it gives the ordinary connected
components).
Indeed they cannot be larger because of the factor
$\prod_{l \in \gF_{1}}  \ep^{\Ga}_{l}$, nor can they be smaller
because of the factor
$\prod _{l \not\in \gF_{1}}  \bigl(\et^{\Ga}_{l} + \ep^{\Ga}_{l}
 h^{\cF,1}_{l} ({\bf h})\bigr) $
which is zero if there are some generalized
neighbors (for which
$\et^{\Ga}_{ij}=0$)
belonging to different clusters (for which $h^{\gF}_{ij} (h)=0$). Therefore
the first forest in this
formula simply automatically draws connecting trees of ``neighbor links''
connecting each such generalized
connected component, but in a symmetric way without
arbitrary choices. Remark that the factor
$\prod _{l \not\in \gF_{1}}  \bigl(\et^{\Ga}_{l} + \ep^{\Ga}_{l}
 h^{\cF,1}_{l} ({\bf h})\bigr)$ is bounded by one as expected
for further estimates. Then the second forest of the jungle automatically takes
into account the first links, i.e. the existence of large field regions, to
draw propagators connections, In this way the units connected by the full
forest remain unit cubes instead of being either small field
cubes or blocks of large field cubes\footnote*{These blocks lead to
unpleasant additional
sums for where in the blocks functional derivatives really act, etc...}.
For convergence of the thermodynamic limit one has
then simply to check that all connections are summable (this is obvious
for the finite range nature of the
$\ep_{l}^{\Ga}$ links), and that there is a small
factor per link of the forest. For $\ep$ links it comes from
the probabilistic factor associated to the presence
of the function $\ch_{\Ga}$ and for ordinary links, it comes either
from the
functional derivative localized out of $\Ga$, or from the large
distance
(at least $M$) crossed in the case of a link between two large field cubes.
This distance induces a small factor through the decay of $C$.
In conclusion there
is no need to gain a small factor from functional derivatives localized
in the large field region (which is usually impossible anyway), and the
whole convergence becomes much more transparent, many combinatoric
difficulties being hidden in the jungle formula itself.

A concrete example in which this formula would somewhat simplify the argument
is e.g. the single scale expansion of [L]; in [KMR] a three level jungle
formula is used, in which the third forest hooks some cubes
along the straight paths of the second forest propagators, in order
to complete factorization in the context of a Peierls contour argument.

\bigskip
\noindent{\bf F) Cluster or resolvent expansions with smooth localizations}
\medskip
In some situations (for instance if momentum
conservation is important), it may be inconvenient to
perform cluster expansions with sharp localization functions
such as $\ch_{b}$ in (V.A.8). But with smooth functions there is
no na\"\i ve factorization in (V.A.8). This is not a serious difficulty and it
can be overcome e.g. by an auxiliary expansion
(sometimes called a ``painting expansion'') on the interaction
that creates protection belts around the clusters.
But in this last section
we remark that the Taylor forest formula (III.2) also gives
elegant solutions to this type of problems and treat the
$\ph^{4}$ interaction again as an example.

Let
$$
1=\sum_b\chi_b^{2} (x)\eqno({\rm V.F.1})
$$
be a smooth partition of unity of $\bbbr^{d}$ by
cubes of unit size.
We assume that the support of $\chi_b$ is contained in
$\{x | {\rm dist}(x,b)\le 1/3 \}$. The
set of all $b$'s is then further restricted to the finite set
$\cN$ of all cubes which are at distance zero of
our finite volume $\La$ (hence include a unit corridor around it).
Therefore
$$
\ch_{\La}(x) =\ch_{\La}(x)\sum_{b \in \cN}
\chi_b^{2}(x)\eqno({\rm V.F.2})
$$
so that the corresponding sums are finite from now on.

Let us rewrite the $\ph^{4}$ theory of section V.A in terms of
an intermediate ultralocal field:

$$
Z(\La)= \int d\mu_{C} (\ph )e^{-g\int _{\La}\ph^{4}(x)dx}=
\int d\mu_{C_{\La}} (\ph ) d\nu (V) e^{i\sqrt g\int _{\La} \ph^{2}(x)V(x)dx}
\eqno({\rm V.F.3})
$$
where $d\nu$ is the normalized Gaussian measure
with ultralocal covariance $\de(x-y)$.

Inserting the identity V.F.2 we get
$$
Z(\La)=
\int d\mu_{C} (\ph ) d\nu (V) e^{i\sqrt g\int_{\La} dx
\sum_{b \in \cN}(\ph\ch_{b})^{2}(x)
V(x)}
\eqno({\rm V.F.4})
$$
\noindent
We define now the collection of Gaussian random variables
$\{\ph_{b} (x)\}$, $b\in \cN$ distributed according to
the measure $d\mu(\{\ph_{b}\} )$ with covariance
$C(b, x~; b', y)=\ch_{b}(x) C(x,y) \ch_{b'}(y)$, and the collection of
Gaussian random variables $\{V_{b}(x)\}$ distributed with
degenerate covariance
$\Ga (b, x~; b', y)=\de (x-y)$. We have
$$
Z(\La)=
\int d\mu (\{\ph_{b}\} ) d\nu (\{V_{b}\}) e^{i\sqrt g\int_{\La} dx
\sum_{b \in \cN}(\ph_{b})^{2}(x)
V_{b}(x)}
\eqno({\rm V.F.5})
$$
(to prove (V.F.5), remark that the right hand side of (V.F.4) and (V.F.5)
are both Borel summable functions of $g$ with identical
perturbative series). We may now change slightly the boundary condition,
dropping the restriction on the interaction range of integration.
In other words $Z(\La) $ leads to the same thermodynamic variables
than
$$
\bar Z(\La)=
\int d\mu (\{\ph_{b}\} ) d\nu (\{V_{b}\}) e^{i\sqrt g
\sum_{b \in \cN}\int dx (\ph_{b})^{2}(x)
V_{b}(x)}
\eqno({\rm V.F.6})
$$

Let us apply the
Taylor forest formula (III.2) to $\bar Z(\La)$,
interpolating both
the covariances $C(b, x~; b', y)$ and $\Ga(b, x~; b', y)$, viewed
as finite matrices with entries in $\cN$
whose elements are operators on $L^{2}(\RR^{d})$.

It gives a result very similar to (V.A.7-8):

$$
\bar Z(\La)=
 \sum_{Y_{1},...,Y_{n} \atop
Y_{i} \cap Y_{j} = \emptyset, \cup Y_{i} =\cN} \prod_{i=1}^{n} A(Y_{i})
$$
$$
A(Y) =
\sum_{T \ {\rm on }\ Y} \bigl(
\prod_{l\in T}\int_{0}^{1} dh_{l} \bigr)
\int d\mu_{C_{Y}^{T}({\bf h}) }(\ph ) d\nu_{\Ga_{Y}^{T}({\bf h}) }(V )
$$
$$\biggl\{\prod_{l = \{ij \} \in T} \int dx dy \biggl( \ch_{b_{i}}(x)
\ch_{b_{j}}(y)C(x,y)
 {\de \over \de \ph_{b_{i}} (x) }{\de \over \de \ph_{b_{j}} (y) } + \de(x-y)
{\de \over \de V_{b_{i}} (x) }{\de \over \de  V_{b_{j}}(y) } \biggr)
\biggr\} $$
$$ e^{i\sqrt g
\sum_{b \in Y}\int dx (\ph_{b})^{2}(x)
V_{b}(x)}
\eqno({\rm V.F.7})
$$
where $b_{i}$ and $b_{j}$ are the two ends of the line $l$, and the sum
is over trees $T$ which connect together the set $Y$, hence have exactly
$|Y|-1$ elements. The measures $ d\mu_{C_{Y}^{T}({\bf h}) }(\ph)$
and $ d\nu_{\Ga_{Y}^{T}({\bf h}) }(V)$
are normalized Gaussian measure on the restricted collections of
Gaussian random variables $\{\ph_{b}(x)\}$ and $\{V_{b}(x)\}$ for
$b \in Y$. $ d\mu_{C_{Y}^{T}({\bf h}) }$
has covariance
$C_{Y}^{T}({\bf h}) (b, x~; b', y) = C(x,y)h_{T}(b,b') $,
where
$h^{T}(b,b)= 1$ and for
$b \ne b'$, $h^{T}(b,b')$
is the infimum of all parameters $h_{l}$ for $l$ in the unique path in
$T$ joining the cube $b$ to the cube
$b'$.
Similarly  $ d\nu_{\Ga_{Y}^{T}({\bf h}) }(V)$ has covariance
$\Ga_{Y}^{T}(b, x~; b', y)=\de(x-y)h_{T}(b,b') $.

Remark that the difference between (V.A.8) and (V.F.7) appears
in the term $\de(x-y)
{\de \over \de V_{b_{i}} (x) }{\de \over \de  V_{b_{j}}(y) }$
which would be zero if the functions $\ch_{b}$ were sharp characteristic
functions. This term corresponds to ``emptying'' the interaction on the borders
of the clusters created so that full factorization occurs, even without
sharp localization functions.

Similar formulas can be written for other kind of interactions or for
resolvent expansions with smooth localization functions.
We leave them to the reader.

\bigskip
\noindent{\bf Acknowledgements}
\medskip

We thank D. Brydges for introducing us to his formula
during the Vancouver 1993 summer school; it is a pleasure
to update our Ecole Polytechnique software accordingly.
We thank also J. Magnen and H. Kn\"orrer for their help and
for discussions on the
proof of all these formulas.

\eject
\noindent{\bf REFERENCES}
\medskip

\item{[BaF]} {G. A. Battle and P. Federbush,  A phase cell cluster
expansion for Euclidean field theories, Ann. Phys. 142, 95 (1982)~;
A note on cluster expansions,
tree graph identities, extra 1/N! factors!!! Lett. Math. Phys. 8, 55
(1984).}

\item{[Bat]}{G. Battle, A new combinatoric estimate for cluster expansions,
Comm. Math. Phys. 94, 133 (1984).}

\item{[B1]} {D. Brydges, A short course on cluster expansions, in Critical
phenomena, random systems, gauge theories, Les Houches session XLIII, 1984,
Elsevier Science Publishers, 1986.}
\item{[B2]} {D. Brydges, Functional integrals and their applications,
``Cours de Troisi\`eme Cycle de la Physique en Suisse Romande''
taken by R. Fernandez, Universit\'e
de Lausanne or University of Virginia preprint, (1992).}

\item{[BF]} {D. Brydges and P. Federbush,
A new form of the Mayer expansion in
classical statistical mechanics, Journ. Math. Phys. 19, 2064 (1978).}

\item{[BK]}{D. Brydges, T. Kennedy, Mayer expansions and the Hamilton-Jacobi
Equation, Journ. Stat. Phys.  48, 19 (1987).}

\item{[BY]} {D. Brydges and H.T. Yau,  Grad $\Ph$ Perturbations of Massless
Gaussian Fields, Comm. Math. Phys.  129, 351 (1990)}

\item{[GJS1]} {J. Glimm, A. Jaffe and T. Spencer,
The Wightman axioms and particle structure in the
$P(\phi)_{2}$ quantum field model, Ann. Math. 100, 585 (1974).}

\item{[GJS2]} {J. Glimm, A. Jaffe and T. Spencer,
The particle structure of the
weakly coupled $P(\phi)_{2}$ model and other applications of high temperature
expansions, Part II: The cluster expansion, in
Constructive Quantum field theory, ed. by G. Velo and A. Wightman,
Lecture Notes in Physics, Vol. 25, Springer (1973).}

\item{[KMR]}{C. Kopper, J. Magnen and V.Rivasseau,
Mass Generation in the two dimensional
Gross-Neveu Model, in preparation.}

\item{[R1]}{V.Rivasseau, From perturbative to constructive renormalization,
Princeton University Press (1991).}

\item{[R2]}{V.Rivasseau, Cluster Expansions with small/large field conditions,
preprint Ecole polytechnique, to appear in Proceedings of the
Vancouver Summer School(1993).}

\end